\documentstyle[11pt,emulateapj]{article}

\newcommand{\Htwo}{H$_2$\ }

\newcommand{\HST}{{\it HST}\ }
\newcommand{\FUSE}{{\it FUSE}\ }
\newcommand{\kms}{km~s$^{-1}$\ }
\newcommand\etal{et~al.\ }

\begin{document}
\title{HIGHLY IONIZED HIGH-VELOCITY CLOUDS: HOT INTERGALACTIC MEDIUM
 OR GALACTIC HALO?}
\author{Joseph A. Collins, J. Michael Shull\altaffilmark{1}}
\affil{University of Colorado, CASA, Department of Astrophysical \&
     Planetary Sciences, Campus Box 389, Boulder, CO 80309 \\
    jcollins@casa.colorado.edu, mshull@casa.colorado.edu}
\altaffiltext{1}{Also at JILA, University of Colorado and National
     Institute of Standards and Technology.}
\and
\author{Mark L. Giroux}
\affil{East Tennessee State University, Department of Physics, Astronomy,
   \& Geology, \\
    Box 70652, Johnson City, TN 37614 \\ giroux@polar.etsu.edu}

\begin{abstract}

We use spectroscopic data from \HST and \FUSE to study the wide range of
ionization states of the ``highly ionized high-velocity clouds".  Studied
extensively in O~VI absorption, these clouds are usually assumed to be
infalling gas in the Galactic halo at distances less than 50 kpc.  
An alternative model attributes the O~VI (and O~VII X-ray absorption) to 
cosmological structures of low-density, shock-heated intergalactic gas,
distributed over 1--3 Mpc surrounding the Milky Way.  The latter 
interpretation is unlikely, owing to the enormous required mass of gas 
($4 \times 10^{12}~M_{\odot}$).  Our detection, in 9 of 12 sight lines, 
of low ionization stages (C~II/III/IV; Si~II/III/IV) at similar high 
velocities as O~VI requires gas densities far above that
($n_H \approx 5 \times 10^{-6}$~cm$^{-3}$) associated with the warm-hot
intergalactic medium (WHIM).  These HVCs are probably cooling, multiphase 
gas in the Galactic halo, bow-shocks and interfaces between clouds falling 
through a hot, rotating gaseous halo.  The velocity segregation of these
HVCs in Galactic coordinates is consistent with a pattern in which
infalling clouds reflect the sense of Galactic rotation, with peculiar
velocities superposed.

\end{abstract}


\section{INTRODUCTION}

This paper deals with a select population of highly ionized Galactic
high-velocity clouds (HVCs), originally discovered in C~IV (Sembach \etal 1999)
and studied in considerable detail in O~VI and other ions (Sembach \etal 2003;
Collins, Shull, \& Giroux 2004 -- hereafter
denoted S03 and CSG04). Most observers assume that this high-velocity
O~VI is associated with infalling clouds in the Galactic halo, including
well-known objects such as the Magellanic Stream and Complex C.  However,
in an alternative interpretation, Nicastro \etal (2002, 2003) suggest that
the high-velocity O~VI, together with the $z=0$ X-ray absorption from O~VII,
arise in the warm-hot intergalactic medium (WHIM) surrounding the Galaxy.  
In their model, the O~VI and O~VII co-exist in a filament of shock-heated 
intergalactic medium (IGM), distributed over 1--3 Mpc from the Local Group 
at very low hydrogen densities, $n_H \approx 5 \times 10^{-6}$~cm$^{-3}$, 
and high temperatures, $T \approx 10^{5.7-6.5}$~K.

The HVCs possess velocities incompatible with differential Galactic
rotation (Wakker \& van Woerden 1997) and are normally detected
through their \ion{H}{1} 21-cm emission.  They display a variety of
different morphologies, ranging from large cloud complexes such as HVC
Complex C (Wakker \etal 1999; Gibson \etal 2001; Richter \etal 2001;
Collins \etal 2003; Tripp \etal 2003) to objects with small angular
sizes such as the compact HVCs (CHVCs; Putman \etal 2002).  The recent
{\it Far Ultraviolet Spectroscopic Explorer} (\FUSE) survey (S03) of
\ion{O}{6} reveals that high-velocity gas can be detected in
\ion{O}{6} resonance absorption lines in $60-85$\% of quasar sight
lines.  Although some of the \ion{O}{6} correlates with
\ion{H}{1}-detected HVCs with column densities
$N$(\ion{H}{1})~$\geq10^{18.5}$ cm$^{-2}$, the S03 survey catalogs a
new class of HVCs with \ion{H}{1} column densities below the detection
threshold of single-dish radio telescopes.  Following the convention
established in early studies of such objects, we refer to the
\ion{O}{6} HVCs without \ion{H}{1} 21-cm detections as the ``highly
ionized HVCs".  Sembach \etal (1995, 1999) analyzed {\it HST}-Goddard
High Resolution Spectrograph (GHRS) data and found that these
absorbers are characterized by strong \ion{C}{4} absorption,
accompanied by little or no absorption in low ions, hence the term
``highly ionized."  Studies of the \ion{O}{6} HVCs have recently been
reviewed by Sembach (2004).

As is the case with most of the larger population of HVCs, the origin
of the highly ionized HVCs is uncertain because their distances are
unknown.  Although the more extended HVC complexes must reside in or
near the Galactic halo ($d<30$ kpc), one theory posits that the
smaller-angular-size CHVCs could be located at large distances
($d\sim1$ Mpc) within the Local Group (Blitz \etal 1999; Braun \&
Burton 1999).  Since the highly ionized HVCs are isolated absorbers,
unconnected to extended objects mapped in \ion{H}{1}, it is possible
that such objects share a similar distribution as proposed for the
CHVCs within the Local Group.  In fact, when the absorption in these
objects in species ionized to \ion{C}{4} and below has been modeled
with single-phase photoionization models, the inferred densities
($n_{H}\sim10^{-4}$ cm$^{-3}$) and pressures ($P/k=1-10$ cm$^{-3}$~K) 
suggest a Local Group location (Sembach \etal 1999; CSG04).  However,
the simultaneous \ion{O}{6} and \ion{C}{4} detections in HVCs imply a
significant collisionally ionized component, arising from immersion in
a surrounding hotter medium (CSG04; Fox \etal 2004, hereafter F04).  
When we attribute both the \ion{O}{6} and \ion{C}{4} absorption 
in the highly-ionized HVCs to the collisionally ionized component, the 
photoionized component can be considerably denser 
($n_{H}>10^{-4}$ cm$^{-3}$), with pressures more indicative of a
Galactic halo location in which $P/k>50$ cm$^{-3}$~K.

In contrast, Nicastro \etal (2002, 2003) suggest that the high-velocity 
\ion{O}{6} absorbers trace large-scale (Local Group) filaments of the 
low-density ($n_H \approx 5\times10^{-6}$ cm$^{-3}$) WHIM.
Cosmological simulations predict that a large fraction of the baryons
in the low-redshift IGM are in the form of shock-heated
($T\sim10^{5}-10^{7}$ K) low-density structures (Cen \& Ostriker 1999;
Dav\'e \etal 1999).  The Nicastro \etal (2003) model for the $z=0$ absorbers 
is based on an interpretation of two observations.  The first comes from
the kinematics of the high-velocity \ion{O}{6} absorbers, which
suggest a preferred reference frame relative to the Local Group
barycenter instead of the Galaxy.  We discuss this issue in depth in
\S~4.1 (see also Fig.\ 1 for the locations and redshifts/blueshifts of
the highly-ionized HVC sight lines, including the 12 we have studied).
The second important observation is the detection of poorly-resolved
\ion{O}{7} and \ion{O}{8} X-ray absorption lines at $z=0$ towards
PKS~2155$-$304, 3C 273, H 1821+643, and Mrk 421
(Fang \etal 2002; Fang, Sembach, \& Canizares 2003; Mathur, Weinberg, \&
Chen 2003; Nicastro \etal 2004) and the speculation that this
absorption can be associated with the \ion{O}{6} HVCs in those sight lines.
McKernan, Yaqoob, \& Reynolds (2004) recently discussed O~VII absorption 
at $z=0$ towards 15 AGN.  
 
\begin{figure*}
\figurenum{1}
\epsscale{1.25}
\plotone{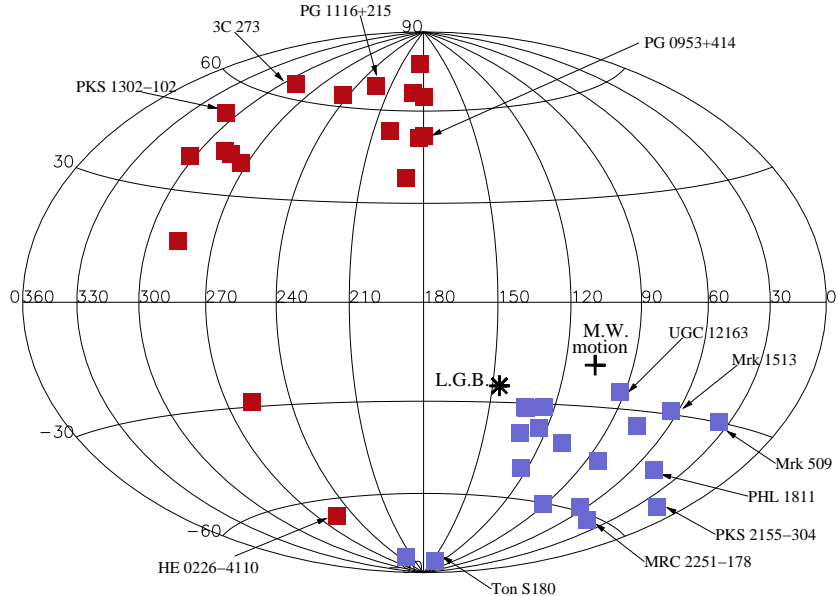}
\caption{All-sky Hammer-Aitoff projection in Galactic coordinates of the
locations of 36 QSO sight lines containing highly ionized HVCs (S03).
The 12 sightlines studied here and in CSG04 are labeled.
Sightlines toward known H~I (21 cm) structures are not shown, which accounts
for the paucity of structure in upper-right and lower-left portions of plot.
These HVCs are not detected in \ion{H}{1} and are designated as
``Local Group'' (PKS 2155$-$304, Mrk 509, MRC 2251$-$178, PHL 1811, Ton S180), 
``extreme positive (north)'' (PG 0953+414, PG 1116+215, 3C 273, 
PKS 1302$-$102), or ``Magellanic Stream
extension'' (Mrk 1513, UGC 12163, HE 0226$-$4110) in the S03 survey.
Positive-velocity HVCs are colored red, while negative-velocity HVCs are blue.
We mark the Local Group barycenter ($\ell = 147\arcdeg$, $b = -25\arcdeg$;
Karachentsev \& Makarov 1996) by an asterisk, and note with a plus sign
the direction of the Milky Way motion ($V = 90$ km~s$^{-1}$, $\ell = 107\arcdeg$,
$b = -18\arcdeg$; Einasto \& Lynden-Bell 1982) with respect to the Local Group
barycenter. }
\end{figure*}

A number of important astrophysical issues are at stake in these two models,
which for simplicity we classify as the WHIM and the Galactic halo. Among 
these issues are: (1) the total mass in
HVCs, which depends on the square of the assumed distance; (2) the role
of HVCs in Galaxy assembly, gaseous infall, and chemical evolution; and 
(3) the possible connection between UV tracers of $10^{5-6}$~K gas (O~VI, 
Ne~VIII) and X-ray tracers of much hotter gas at $10^{6-7}$~K (O~VII, O~VIII, 
etc).  The last issue could provide a connection between O~VI studies of Galactic
HVCs and cosmological measurements of the baryon content in the multiphase
IGM at low redshift (Shull 2003;  Nicastro \etal 2003; Stocke, Shull, \&
Penton 2004).

The Nicastro \etal (2002, 2003) conclusions have been disputed (CSG04;
Sembach 2004), since many \ion{O}{6} HVCs are correlated in velocity
and space with known Galactic-halo HVCs mapped in \ion{H}{1} 21 cm
emission.  More significantly, \HST and \FUSE detections (CSG04) of
low ions at the same velocities as O~VI require gas densities much
higher than those in the WHIM. For example, the existence of
\ion{C}{2} and \ion{Si}{2} in the highly ionized HVCs towards
PKS~2155$-$304 suggests a characteristic hydrogen density $n_H >
10^{-4}$ cm$^{-3}$, and models of the HVC Complex C (Collins, Shull,
\& Giroux 2003, hereafter CSG03) suggest $n_H =$
0.01--0.1 cm$^{-3}$.  CSG04 argue that the data are best explained if
the PKS~2155$-$304 highly ionized HVCs are low total-hydrogen column
density analogs of the Galactic halo HVCs detected in 21 cm emission,
rather than diffuse WHIM filaments.

In order to assess possible origins for the highly ionized HVCs, it is
important to consider the population as a whole, instead of focusing on
individual sight lines.  We consider the statistics of a sample of 12
highly ionized HVCs, none of which is directly associated with known
nearby 21 cm structures (S03). From these sight lines,
we can better investigate whether the population traces the WHIM, or
instead is a low-column version of the Galactic halo HVCs.  Our previous 
study (CSG04) investigated highly ionized HVCs in two sight
lines, PKS 2155$-$304 and Mrk 509, and found low-ionization absorbers in both
cases.  In this paper, we present \FUSE and \HST Space Telescope Imaging
Spectrograph (STIS) data for ten more sight lines containing highly ionized
HVCs detected in the S03 survey.  We have now analyzed 12 such sight lines, 
nine of which show low-ionization species at the same velocity as \ion{O}{6}.

The \FUSE mission and instrument are discussed in Moos \etal (2000) and
Sahnow \etal (2000). In \S~2, we present the sight-line selection criteria and
data-reduction procedures.  The ion species detections and column density
measurements for each sight line are reported in \S~3.  In \S~4, we discuss the
results and investigate the ionization pattern in the HVCs via photoionization
and collisional ionization models. We present our conclusions in \S~5.

\section{SIGHT LINE SELECTION AND DATA}

The ten sight lines selected for this work were chosen from the \FUSE survey
of high-velocity \ion{O}{6}\ absorption (S03).  In this work, we include all
sight lines from that survey with good quality STIS E140M and G140M data
that are publicly available as of early April 2004.  Using this criterion, 
we can analyze a variety of resonance lines beyond those available in the \FUSE
data alone.  Absorption lines of \ion{O}{1}, \ion {C}{2}, and \ion{Si}{2}
in the STIS bandpass are typically much
stronger\footnotemark\footnotetext{The strength of a resonance line is
quantified by the product, $f\lambda$, of oscillator strength and wavelength
(Morton 2003).} than lines in the \FUSE bandpass, and thus are more sensitive
for measurement of the low column densities typical of highly ionized HVCs.
We include only the sight lines with \FUSE data quality of $S/N>5$ per
20 \kms resolution element (those with a data quality factor
$Q \geq 2$ in Table 1 of S03).  This requirement eliminates only one sight line,
Mrk 926, from this study.  For Mrk 926, we detect high-velocity \ion{Si}{3}
in the STIS G140M data associated with the reported high-velocity \ion{O}{6}
detection from S03.  However, we are unable to measure $N$(\ion{O}{6}) of this
absorber above the 3$\sigma$ threshold.  A summary of the \FUSE and \HST
observations is given in Table 1.

\begin{deluxetable}{lllllll}
\tablefontsize{\footnotesize}
\tablecolumns{6}
\tablewidth{0pc}
\tablecaption{Summary of Observations \label{t1}}
\tablehead{
\colhead{} & \colhead{FUSE} & \colhead{FUSE} & \colhead{\ \ } &
   \colhead{HST-STIS} & \colhead{HST} & \colhead{HST-STIS}\\
\colhead{Sightline} & \colhead{Program ID} & \colhead{$T_{\rm exp}$(ks)\tablenotemark
{a}} & \colhead{\ } &
\colhead{Grating} & \colhead{Program ID} & \colhead{$T_{\rm exp}$(ks)}}
\startdata
3C 273       & P101       & 43.2  & \ \  & E140M  & 8017   & 18.7 \\
HE 0226-4110 & P101, D027 & 192.6 & \ \  & E140M  & 9184   & 43.8 \\
MRC 2251-178 & P111       & 51.4  & \ \  & G140M  & 7345   & 6.0,
    4.6\tablenotemark{b} \\
Mrk 1513     & P101       & 22.5  & \ \  & G140M  & 7345   & 6.2,
    7.3\tablenotemark{b}\\
PG 0953+414  & P101       & 74.4  & \ \  & E140M  & 7747   & 24.5 \\
PG 1116+215  & P101       & 76.7  & \ \  & E140M  & 8097, 8165 & 39.9     \\
             &            &       & \ \  & E230M  & 8097   & 5.6      \\
PHL 1811     & P108, P207 & 75.7  & \ \  & E140M  & 9418   & 18.4   \\
             &            &       & \ \  & G230MB & 9128   & 1.2   \\
PKS 1302-102 & P108       & 145.9 & \ \  & E140M  & 8306   & 22.1  \\
Ton S180     & P101       & 16.9  & \ \  & G140M  & 7345   & 4.1,
    3.5\tablenotemark{b}\\
             &            &       & \ \  & G230MB & 9128   & 1.2  \\
UGC 12163    & B062       & 60.9  & \ \  & E140M  & 8265   & 10.3
\enddata
\tablenotetext{a}{Effective exposure time for the LiF1a channel.}
\tablenotetext{b}{The G140M observations are taken at two separate grating tilts. The
 two listed exposure times are for central wavelengths of 1222~\AA\ and 1272~\AA,
 respectively.}
\end{deluxetable}

Calibrated \FUSE spectra were extracted by passing raw data
through the CALFUSE Version 2.4 reduction pipeline.  In order to
improve the signal-to-noise ratio (S/N), both ``day'' and ``night''
photons were included in the final calibrated \FUSE spectra.  The
inclusion of day photons leads to strong airglow contamination of
certain neutral interstellar lines near the local standard of rest
(LSR).  Such airglow contamination is irrelevant for the purpose of
investigating high-velocity absorption in this study.
Individual exposures were
then co-added and weighted by their exposure times, to yield a final
\FUSE spectrum.  The \FUSE spectra have a resolution over the
\FUSE bandpass (912-1187 \AA) of about 20 \kms ($\sim10$
pixels), and as a result the data are oversampled at that resolution.
In order to further improve S/N, the data were rebinned over 5
pixels.  To set the absolute wavelength scale, the centroid of Galactic
\ion{H}{1}\ 21-cm emission was compared and aligned to various
Galactic absorption lines in the \FUSE bandpass, such as
\ion{Si}{2}\ ($\lambda$1020.70), \ion{O}{1}\ ($\lambda$1039.23),
\ion{Ar}{1}\ ($\lambda 1048.22$, $\lambda 1066.66$), \ion{Fe}{2}\
($\lambda 1125.45$, $\lambda 1144.94$), \ion{N}{1}\ ($\lambda$1134.17),
and various H$_{2}$ Lyman bands.

The \HST STIS data consist of observations taken with several different
gratings.  The E140M echelle mode, which provides 7 \kms resolution,
is the preferred grating for this work owing to its bandpass,
1150 -- 1700 \AA, which covers resonance lines of several important ion
species.  Seven of the ten sight lines in this survey were observed in the
E140M mode.  The PG 1116+215 sight line was observed in both the E140M and
E230M echelle modes.  The E230M observations of this sight line provide
10 \kms resolution over 2010 -- 2820 \AA, which covers several
\ion{Fe}{2} and \ion{Mg}{2} lines.  The rest of the data consist of
observations taken with the first-order G140M grating at $\sim$25 \kms
resolution.  The G140M data were all taken for Program 7345, which investigated
the low-$z$ Ly$\alpha$ forest in sight lines of quasars at $z<0.07$ 
(Penton, Stocke, \& Shull 2004).
In each case, the G140M observations were taken at two separate tilts,
resulting in a wavelength coverage, 1195 -- 1299 \AA, which includes
resonance lines of \ion{N}{1}, \ion{Si}{3}, \ion{N}{5}, \ion{S}{2}, and
\ion{Si}{2}.  In two cases, we supplement the E140M and G140M
data with observations taken with the G230MB grating, covering 
$2760-2910$ \AA\, including the \ion{Mg}{2} doublet at 2800 \AA\, at a 
resolution of $\sim30$ km~s$^{-1}$.  Final STIS spectra were
obtained by co-adding individual exposures, weighted by their exposure times.
The E140M and E230M echelle data were rebinned over 3 pixels, while no
rebinning was used for the first-order G140M and G230MB spectra.
Absolute wavelength scales were
obtained as for the \FUSE spectra, matching the Galactic
\ion{H}{1}\ emission peak to absorption features of \ion{N}{1}\
($\lambda$1199.55, 1200.71), \ion{S}{2}\ ($\lambda$1250.58, 1253.80, 1259.52),
\ion{Fe}{2}\ ($\lambda$1608.45, 2374.46, 2586.65), and \ion{Mg}{2}\
($\lambda\lambda$2796.35, 2803.53). We estimate that, with our technique
of registering interstellar lines to 21 cm emission, the absolute wavelength
scales of both the STIS and \FUSE data are accurate to $\sim10$ \kms
(Penton \etal 2000; Indebetouw \& Shull 2004; Savage \etal 2005).

\section{COLUMN DENSITY MEASUREMENTS}

In the following subsections, we discuss results for HVCs in the ten sight 
lines considered in this survey. 
Since the goal of this study is to ascertain the physical conditions
of these absorbers and their environments, it is crucial to determine
column densities of ion species using UV resonance lines.  From the
full \FUSE and \HST STIS spectra, we extract individual line
profiles and normalize them by fitting low-order polynomials to the
continuum immediately surrounding the line in question.  Typically, we
fit the continuum $\pm3-5$ \AA\ about the rest wavelength of the line,
although in a number of cases spurious absorption near the line
required the use of a much larger wavelength band for continuum measurement.
There are many useful resonance lines in the \FUSE and \HST STIS
bandpass.  However, given the column densities typical of the highly-ionized
HVCs and the sensitivity of the observations, the ion species
that could provide reliable constraints on the HVC characteristics include
\ion{O}{6}, \ion{N}{5}, \ion{C}{4}, \ion{Si}{4}, \ion{Si}{3}, \ion{C}{3},
\ion{N}{3}, \ion{C}{2}, \ion{Si}{2}, \ion{Fe}{2}, \ion{Mg}{2}, \ion{N}{2}, 
and \ion{O}{1}.
We have not detected any \ion{N}{5}, which is typically weak relative
to the other high ions.

Because the HVC line profiles typically become broader at higher
ionization states, we generally use the velocity extent of the high-velocity
\ion{O}{6} to establish the HVC integration range.
We use the apparent optical depth (AOD) method, valid for unsaturated lines
(Savage \& Sembach 1991), to measure column densities from the line profiles.
Line saturation can be difficult to detect,
particularly if such saturation is unresolved.  For these absorbers,
most of the profiles are consistent with the optically thin limit for
the AOD method.  The exceptions are the strong lines of
\ion{C}{3} $\lambda977.02$ and \ion{Si}{3} $\lambda1206.50$, which are
frequently saturated.  In those cases, resulting AOD column density
measurements are taken as lower limits.  We have taken care 
that none of the measured absorption features include IGM 
contamination.  In several cases, the negative high-velocity \ion{O}{6} 
profiles are contaminated by weak absorption from the Lyman
[L(6-0)P(3)] Galactic \Htwo line at 1031.19 \AA.  To remove this
contamination, we determine the optical depth, $\tau_{v}$, of other \Htwo
($J=3$) lines in the \FUSE bandpass and infer the optical depth of 
L(6-0)P(3) based on its relative value of $f\lambda$ (Abgrall \etal
1993a,b).  The line widths of the various $J=3$ lines
are then used to subtract off the appropriate integrated $\tau_{v}$ for the
contaminating absorption.

The \FUSE bandpass covers \ion{H}{1} Lyman lines from Ly$\beta$ down to
the Lyman limit, which can be used to constrain the \ion{H}{1} column density
of the HVC under certain circumstances.  In our previous study (CSG04), the
measured HVC column densities based on Lyman-line absorption were crucial
in constraining photoionization models.  In this work, however, we are unable
to measure \ion{H}{1} Lyman lines for the HVCs, except in the case of
PG~1116+215.  The two factors preventing these measurements include the
common occurrence of poor quality in the SiC channels and blending
of HVC with Galactic \ion{H}{1} absorption.

\subsection{3C~273}

\begin{figure*}
\figurenum{2}
\epsscale{0.7}
\plotone{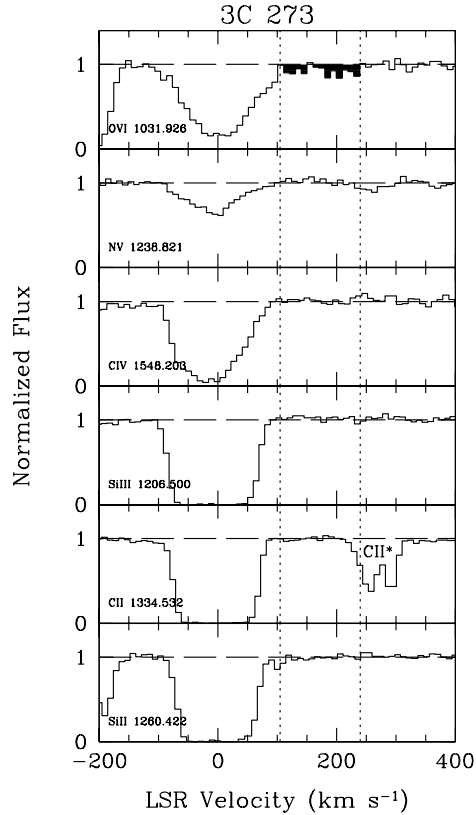}
\caption{Normalized absorption profiles from STIS E140M and {\it FUSE}
data for 3C 273.  The vertical dashed lines indicate the velocity
extent of the high-velocity component determined from its \ion{O}{6} absorption
(105 to 240 \kms).  The shaded region indicates absorption from
the high-velocity component, which in this case can only be detected for the
\ion{O}{6} line.}
\end{figure*}

\begin{deluxetable}{lrccc}
\tablefontsize{\footnotesize}
\tablecolumns{5}
\tablewidth{0pc}
\tablecaption{SUMMARY OF MEASUREMENTS: 3C~273 \label{t2}}
\tablehead{
\colhead{} & \colhead{$\lambda$\tablenotemark{a}} & \colhead{} & \colhead{$W_{\lambda
}$} & \colhead{log $N(X)$\tablenotemark{b}} \\
\colhead{Species} & \colhead{(\AA)} & \colhead{} & \colhead{(m\AA)} &
\colhead{($N$ in cm$^{-2}$)}}
\startdata
\ion{O}{6}  & 1031.926 \ \ & &   $41\pm7$ & $13.54^{+0.07}_{-0.10}$ \\
\ion{N}{5}  & 1238.821 \ \ & &   $<15$    & $<12.83$                \\
\ion{C}{4}  & 1548.203 \ \ & &   $<19$    & $<12.66$                \\
\ion{Si}{3} & 1206.500 \ \ & &   $<16$    & $<11.88$                \\
\ion{C}{2}  & 1334.532 \ \ & &   $<18$    & $<13.00$                \\
\ion{Si}{2} & 1260.422 \ \ & &   $<12$    & $<11.86$
\enddata
\tablenotetext{a}{Wavelengths and oscillator strengths are from Morton
   (2003).}
\tablenotetext{b}{All column densities are calculated
through the apparent optical depth method.
The HVC occupies the velocity
range $105<V_{LSR}<240$ km s$^{-1}$ based on the \ion{O}{6} profile.
All upper limits are 3$\sigma$ levels.}
\end{deluxetable}

Owing to its bright UV flux and flat continuum, 3C 273 is a showcase
spectrum for \FUSE and \HST studies of absorption from
interstellar, intergalactic, and high-velocity gas.  Sembach et
al. (2001) detect a positive-velocity wing to Galactic \ion{O}{6}
absorption, but no corresponding absorption in other ion species in
their \FUSE and \HST GHRS data.  They conclude that the \ion{O}{6} wing 
most likely traces hot gas flowing from the Galactic disk in a Galactic 
fountain.  Since that study, Tripp et al. (2002) presented STIS E140M data
for their study of absorbers within the Virgo Cluster.  Using the
\FUSE and STIS E140M data, we searched for the positive velocity
wing in other absorption lines besides \ion{O}{6}.

Line profiles for several key absorption lines are shown in Figure 2.
The positive velocity wing can be detected only in \ion{O}{6}.  We use
the \ion{O}{6} absorption range $105<V_{\rm LSR}<240$ \kms as
the range of integration to establish 3$\sigma$ upper limits on column
densities involving other species, with results of these measurements
shown in Table 2. It is possible that some of the high-velocity \ion{O}{6}
absorption arises from the \Htwo L(6-0)R(4) line at 1032.35 \AA.  The
spectrum shows strong interstellar \Htwo lines from $J=0-3$ rotational
levels, but no absorption from $J=4$ lines.  Thus, we can be reasonably
confident that nearly all of the absorption wing arises from high-velocity
\ion{O}{6}.  The integration range in the \ion{C}{2} profile is partially
contaminated by the \ion{C}{2}$^{*}$ line.  Since the velocity range
$100<V_{\rm LSR}<210$ \kms shows no high-velocity \ion{C}{2} absorption,
we can say with confidence that the optical depth over $210<V_{\rm LSR}<240$
\kms is entirely from interstellar \ion{C}{2}$^{*}$.  As a result, we
use the measured optical depth over the entire integration range to
establish a conservative upper limit on $N$(\ion{C}{2}).  We could not
measure C~III $\lambda$977 because of contamination from Ly$\gamma$ in
an intergalactic absorber at $z = 0.0053$.

\placetable{t2}

\subsection{HE~0226-4110}

\begin{figure*}
\figurenum{3}
\epsscale{0.67}
\plotone{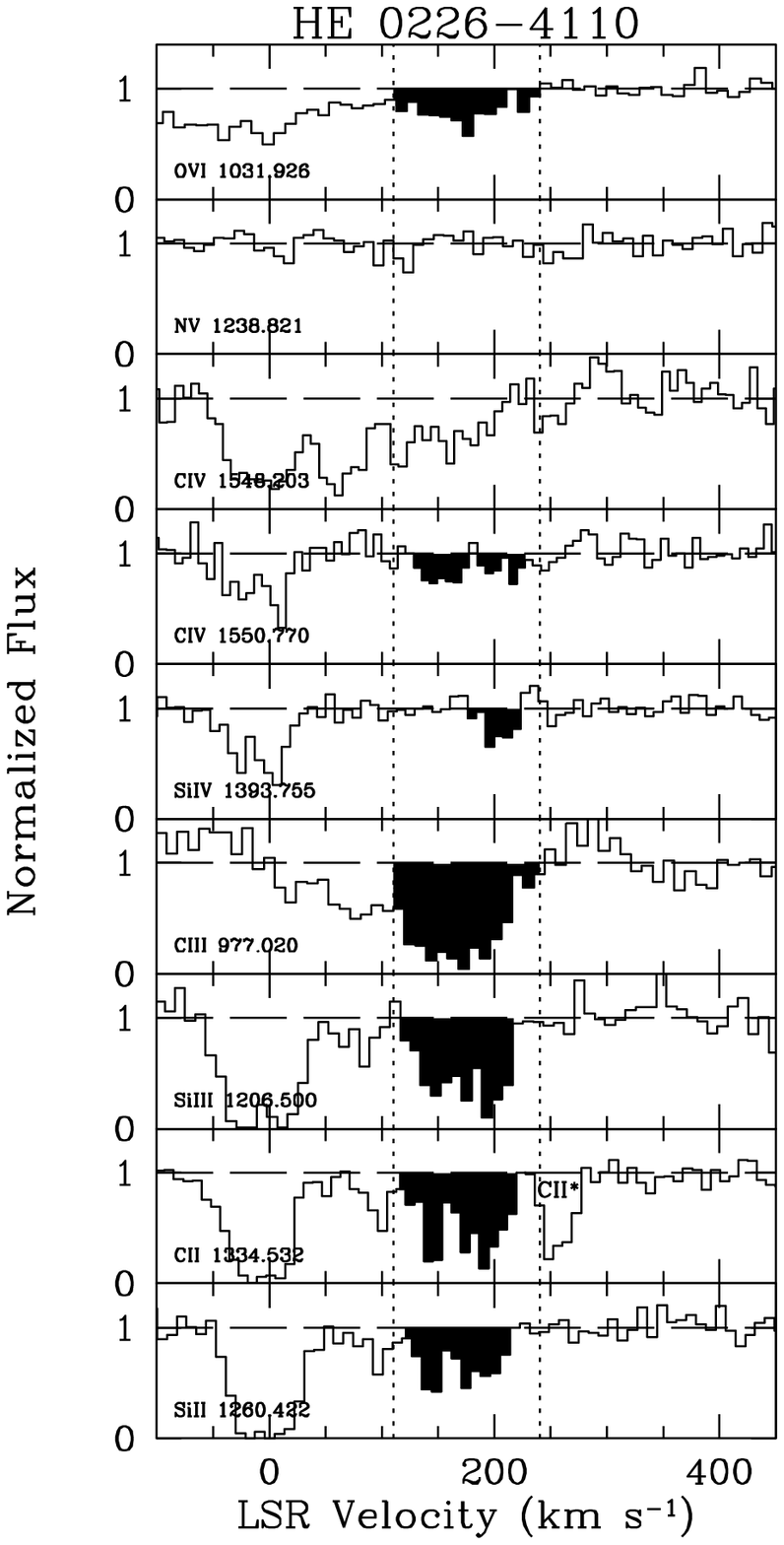}
\caption{Normalized absorption profiles from STIS E140M and {\it FUSE}
data for HE~0226$-$4110.
The shaded regions indicate absorption from the high-velocity component,
while the vertical dashed lines indicate the velocity extent of the
high-velocity \ion{O}{6} absorption ($110$ to $240$ \kms).
The high-velocity component can be detected in lines of
\ion{O}{6}, \ion{C}{4}, \ion{Si}{4}, \ion{C}{3}, \ion{Si}{3}, \ion{C}{2},
and \ion{Si}{2}. The \ion{C}{4}$\lambda1548.203$ profile is contaminated
by \ion{O}{6} intrinsic to the QSO.}
\end{figure*}

\begin{deluxetable}{lrccc}
\tablefontsize{\footnotesize}
\tablecolumns{5}
\tablewidth{0pc}
\tablecaption{SUMMARY OF MEASUREMENTS: HE~0226-4110 \label{t3}}
\tablehead{
\colhead{} & \colhead{$\lambda$\tablenotemark{a}} & \colhead{} & \colhead{$W_{\lambda
}$} & \colhead{log $N(X)$\tablenotemark{b}} \\
\colhead{Species} & \colhead{(\AA)} & \colhead{} & \colhead{(m\AA)} &
\colhead{($N$ in cm$^{-2}$)}}
\startdata
\ \ \ \ion{O}{6} \ \  \ \ \  \ & \ \ \ \ \ 1031.926 \ \  \ \ \ & &  \ \ \ \ \ $84\pm10$ \ \ \ \ \ & \ \ \ $13.89^{+0.04}_{-0.05}$ \ \ \ \\
\ \ \ \ion{N}{5} \ \  \  \ \  \ & \ \ \ \ \ 1238.821 \ \  \ \ \ & &   $<50$    & \ \ \ $<13.38$ \ \ \                \\
\ \ \ \ion{C}{4} \ \  \  \ \  \ & \ \ \ \ \ 1550.770 \ \ \ \  \ & &  $85\pm13$ & \ \ \ $13.68^{+0.06}_{-0.07}$ \ \ \ \\
\ \ \ \ion{Si}{4} \ \  \  \ \  \ & \ \ \ \ \ 1393.755 \ \ \ \  \ & &  $42\pm12$ & \ \ \ $12.74^{+0.10}_{-0.12}$ \ \ \ \\
\ \ \ \ion{C}{3} \ \  \  \ \  \ & \ \ \ \ \ 977.020  \ \ \ \  \ & & $270\pm20$ & \ \ \ $13.94^{+0.03}_{-0.03}$\tablenotemark{c} \ \ \ 
\\
\ \ \ \ion{Si}{3} \ \  \ \ \  \ & \ \ \ \ \ 1206.500 \ \ \ \  \ & & $232\pm18$ & \ \ \ $13.27^{+0.03}_{-0.04}$\tablenotemark{c} \ \ \ 
\\
\ \ \ \ion{C}{2} \ \  \  \ \  \ & \ \ \ \ \ 1334.532 \ \ \ \  \ & & $240\pm11$ & \ \ \ $14.30^{+0.02}_{-0.02}$ \ \ \ \\
\ \ \ \ion{Si}{2} \ \  \ \ \  \ & \ \ \ \ \ 1260.422 \ \ \ \  \ & & $146\pm11$ & \ \ \ $13.06^{+0.03}_{-0.03}$ \ \ \ 
\enddata
\tablenotetext{a}{Wavelengths and oscillator strengths from Morton~(2003).}
\tablenotetext{b}{All column densities are calculated
through the apparent optical depth method.  The HVC occupies the velocity
range $110<V_{LSR}<240$ km s$^{-1}$ based on the various line profiles. 
All upper limits are 3$\sigma$ levels.}
\tablenotetext{c}{Because some saturation is indicated in the line profile, the total
 column density may be slightly higher.}
\end{deluxetable}

High-quality \FUSE and \HST STIS E140M data exist for the
HE~0226$-$4110 sight line.  A very detailed study of high-velocity gas
in this sight line is presented by Fox et al. (2005, in preparation).
Line profiles for several absorption lines (Fig.\ 3) clearly show a
high-velocity component over the velocity range, $110<V_{\rm LSR}<240$
\kms.  There is evidence for a component centered at $V_{\rm LSR} =
80$ km~s$^{-1}$, just below what is formally considered ``high
velocity".  It is detected in \ion{O}{6}, \ion{C}{3}, \ion{Si}{3},
\ion{C}{2}, and \ion{Si}{2}.  Absorption at this velocity in
\ion{C}{4}$\lambda1548.20$ is due to contamination by \ion{O}{6}
absorption intrinsic to the QSO.  Because this feature is probably
blended with low-velocity and Galactic absorption, we concentrate our
effort on the truly high velocity absorption.  The high-velocity
feature is detected in \ion{O}{6}, \ion{C}{4}, \ion{Si}{4},
\ion{C}{3}, \ion{Si}{3}, \ion{C}{2}, and \ion{Si}{2}.  Column density
measurements involving these species are shown in Table 3.  This HVC
is detected in \ion{C}{4}, since high-velocity absorption is detected
over the appropriate velocity range for both lines of the doublet. We
measured C~IV from $\lambda1550.7$ only, since the profile of
$\lambda1548.20$ is contaminated by the AGN intrinsic \ion{O}{6}.  
The \ion{C}{3} profile
shows evidence of saturation, indicating that the measured column
density is likely a lower limit.  While the profile for \ion{Si}{3}
binned to 3 pixels (Fig.\ 3) does not seem saturated, the unbinned
data do suggest that saturation over a narrow velocity range is
likely.  We take the measured $N$(\ion{Si}{3}) as a lower limit for
the modeling in \S~4.  Although the integration range in the
\ion{C}{2} $\lambda1036.34$ profile is badly contaminated by Galactic
\ion{C}{2}$^{*}$ absorption, the 1334.53 \AA\ profile is contaminated
only at $V_{\rm LSR}>230$ \kms.  As a result, the column density of
\ion{C}{2} for the HVC can easily be measured.

\subsection{MRC~2251$-$178}

\begin{figure*} 
\figurenum{4}
\epsscale{0.7}
\plotone{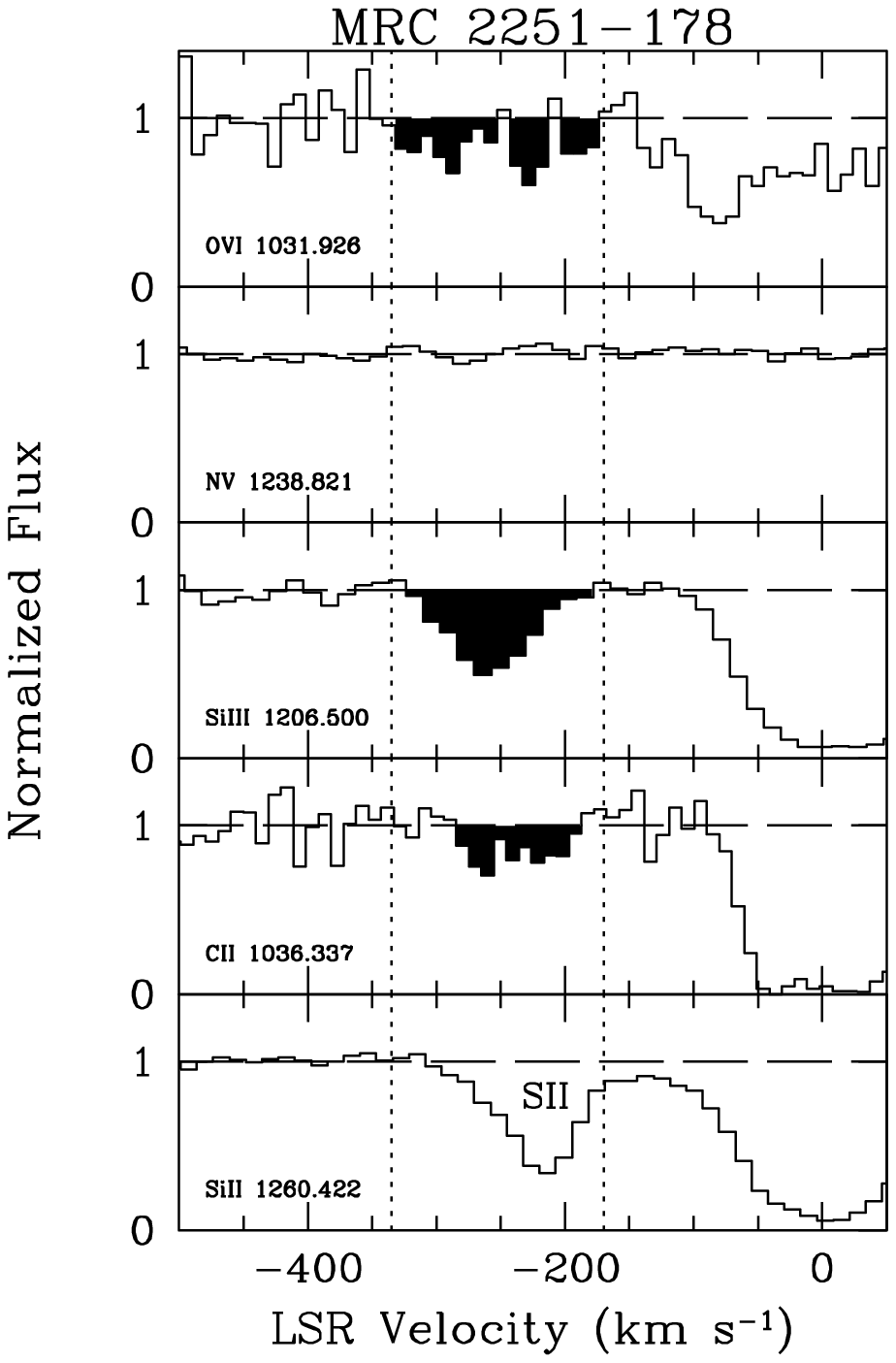}
\caption{Normalized absorption profiles from STIS G140M and {\it FUSE}
data for MRC 2251$-$178.  The shaded regions indicate absorption from the
high-velocity component, while the vertical dashed lines indicate the 
velocity extent of the high-velocity \ion{O}{6} absorption ($-335$ to $-170$ 
\kms).  The high-velocity component can be detected in lines of
\ion{O}{6}, \ion{Si}{3}, and \ion{C}{2}.  We have no \ion{C}{4} or \ion{Si}{4}
coverage for this sight line. }
\end{figure*}

\begin{deluxetable}{lrccc}
\tablefontsize{\footnotesize}
\tablecolumns{5}
\tablewidth{0pc}
\tablecaption{SUMMARY OF MEASUREMENTS: MRC~2251$-$178 \label{t4}}
\tablehead{
\colhead{} & \colhead{$\lambda$\tablenotemark{a}} & \colhead{} & \colhead{$W_{\lambda
}$} & \colhead{log $N(X)$\tablenotemark{b}} \\
\colhead{Species} & \colhead{(\AA)} & \colhead{} & \colhead{(m\AA)} &
\colhead{($N$ in cm$^{-2}$)}}
\startdata
\ion{O}{6}  & 1031.926 \ \ & &  $98\pm27$ & $13.97^{+0.10}_{-0.12}$ \\
\ion{N}{5}  & 1238.821 \ \ & &   $<40$    & $<13.27$                \\
\ion{Si}{3} & 1206.500 \ \ & & $145\pm13$ & $12.93^{+0.03}_{-0.04}$ \\
\ion{C}{2}  & 1036.337 \ \ & &  $52\pm17$ & $13.73^{+0.13}_{-0.17}$ \\
\enddata
\tablenotetext{a}{Wavelengths and oscillator strengths are from Morton (2003).}
\tablenotetext{b}{All column densities are calculated
through the apparent optical depth method.  The HVC occupies the velocity
range $-335<V_{LSR}<-170$ km s$^{-1}$ based on the various line profiles.
All upper limits are 3$\sigma$ levels.}
\end{deluxetable}

The MRC 2251$-$178 sight line has been observed both with \FUSE and the G140M
grating on \HST STIS.  Figure 4 shows line profiles for several absorption
lines.  Two negative-velocity \ion{O}{6} HVCs were identified (S03) in this
sight line, centered at $-258$ and $-95$ km~s$^{-1}$.  The HVC at $-258$ \kms 
is clearly detected in \ion{Si}{3} and \ion{C}{2}.  If the $-95$ \kms trough is
indeed an HVC, it is difficult to assign a minimum
velocity cutoff since the absorption is blended with Galactic
\ion{O}{6}.  Absorption at the same velocity cannot be detected in
\ion{Si}{3}, although weak absorption in the \ion{C}{2} profile may be
attributed to an HVC at such a velocity.  Fluctuations in
the spectrum over 1035--1038 \AA\ make the continuum in this part of
the profile somewhat uncertain.  Since the \ion{O}{6} absorption
cannot be conclusively distinguished from Galactic absorption at
$|V_{\rm LSR}|<100$ \kms, we concentrate our efforts on the HVC at
$-258$ \kms.  Column density measurements for the HVC detected
in \ion{O}{6} over the range $-335<V_{\rm LSR}<-170$ \kms are shown in Table 4.
Although there is H$_{2}$ in the spectrum, we can rule out H$_{2}$
contamination of the \ion{O}{6} and \ion{C}{2} profiles, since absorption
in lines of $J=3$ and $4$ cannot be detected in other parts of the spectrum.
Because of poor data in the \FUSE SiC channels, we cannot properly analyze
the \ion{C}{3} $\lambda977.02$ line.

\placetable{t4}

\subsection{Mrk~1513}

\begin{figure*}
\figurenum{5}
\plotone{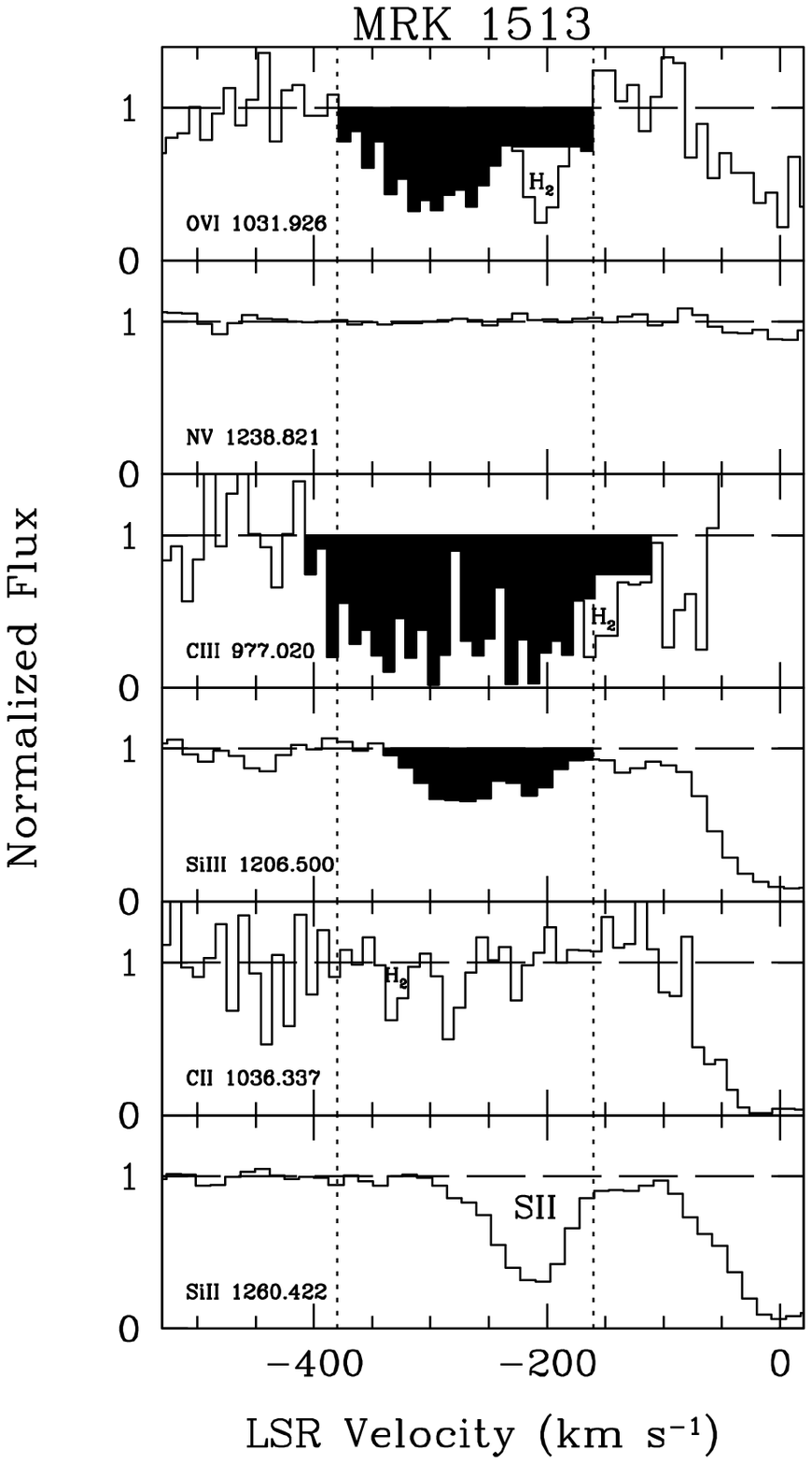}
\caption{Normalized absorption profiles from STIS G140M and {\it FUSE}
data for Mrk 1513.
The shaded regions indicate absorption from the
high-velocity component,
while the vertical dashed lines indicate the velocity extent of the
high-velocity \ion{O}{6} absorption ($-380$ to $-160$ \kms).
The high-velocity component can be detected in lines of
\ion{O}{6}, \ion{Si}{3}, and \ion{C}{3}.  We have no \ion{C}{4} or \ion{Si}{4}
coverage for this sight line.}
\end{figure*}

\begin{deluxetable}{lrccc}
\tablefontsize{\footnotesize}
\tablecolumns{5}
\tablewidth{0pc}
\tablecaption{SUMMARY OF MEASUREMENTS: MRK~1513 \label{t5}}
\tablehead{
\colhead{} & \colhead{$\lambda$\tablenotemark{a}} & \colhead{} & \colhead{$W_{\lambda
}$} & \colhead{log $N(X)$\tablenotemark{b}} \\
\colhead{Species} & \colhead{(\AA)} & \colhead{} & \colhead{(m\AA)} &
\colhead{($N$ in cm$^{-2}$)}}
\startdata
\ion{O}{6}  & 1031.926 \ \ & & $295\pm57$ & $14.51^{+0.07}_{-0.08}$ \\
\ion{N}{5}  & 1238.821 \ \ & &   $<50$    & $<13.37$                \\
\ion{C}{3}  &  977.020 \ \ & & $>380$     & $>14.14$                \\
\ion{Si}{3} & 1206.500 \ \ & & $157\pm15$ & $12.93^{+0.04}_{-0.04}$ \\
\ion{C}{2}  & 1036.337 \ \ & &  $<132$    & $<14.08$                \\
\enddata
\tablenotetext{a}{Wavelengths and oscillator strengths are from Morton (2003).}
\tablenotetext{b}{All column densities are calculated
through the apparent optical depth method.  
The HVC occupies the velocity
range $-380<V_{LSR}<-160$ km s$^{-1}$ based on the various line profiles.
All upper limits are 3$\sigma$ levels.}
\end{deluxetable}

Markarian 1513 has been observed with \FUSE, as well as with 
the STIS G140M grating.  Profiles for several important absorption lines are
shown in Figure 5.  The high-velocity \ion{O}{6} component is strong and
extends over a width of nearly 200 \kms.  The \ion{O}{6} profile
shows evidence of a two-component structure, though S03 identify the
narrow trough centered near $V_{\rm LSR} = -200$ \kms as arising from the
H$_{2}$ L(6-0)P(3) line at 1031.19~\AA.  However, our analysis
of other \Htwo $J=3$ lines in the \FUSE bandpass indicates that some
of that absorption must arise from high-velocity \ion{O}{6}.  The \ion{C}{3}
and \ion{Si}{3} profiles show absorption at this velocity and
seem to confirm the presence of the lower-velocity component.  The absorption
attributed to high-velocity \ion{O}{6} after H$_{2}$ removal
is shown in the shaded regions in Figure 5.

Column density measurements for the HVC detected in \ion{O}{6} over
the range $-380 < V_{\rm LSR} < -160$ \kms are shown in Table 5.  The
\ion{C}{3} profile clearly indicates saturation for the high-velocity
component, which extends beyond the velocity width of the \ion{O}{6}
absorption.  We take the measured column density over the integration
width of the high-velocity \ion{O}{6} as the lower limit on
$N$(\ion{C}{3}).  We do not detect high-velocity absorption in
\ion{C}{2} $\lambda1036.34$.  The feature at $V_{\rm LSR} = -330$ \kms
is likely from H$_{2}$ L(6-0)P(4) at 1035.18~\AA.  The
low-significance absorption at $V_{\rm LSR} = -280$ \kms cannot be
conclusively identified as high-velocity \ion{C}{2}, since the feature
cannot be detected in both LiF channels.  We note, however, that the
\FUSE data for Mrk~1513 are among the worst in our sample.  Comparing to
the results for MRC~2251$-$178, where we measure an identical
high-velocity $N$(\ion{Si}{3}), the data for Mrk 1513 would be of
insufficient quality to measure a high-velocity \ion{C}{2} component
comparable to that in the MRC~2251$-$178 sight line.
The nondetection of high-velocity \ion{C}{2} in this sight line is not
surprising, given the data quality.  The other key single-ion resonance
line, \ion{Si}{2} $\lambda1260.42$, cannot be analyzed, 
owing to severe contamination from Galactic \ion{S}{2} $\lambda1259.52$.
It would certainly help the modeling of this HVC to obtain data
for C~IV and Si~IV.

\placetable{t5}

\subsection{PG~0953+414}

\begin{figure*}
\figurenum{6}
\epsscale{0.65}
\plotone{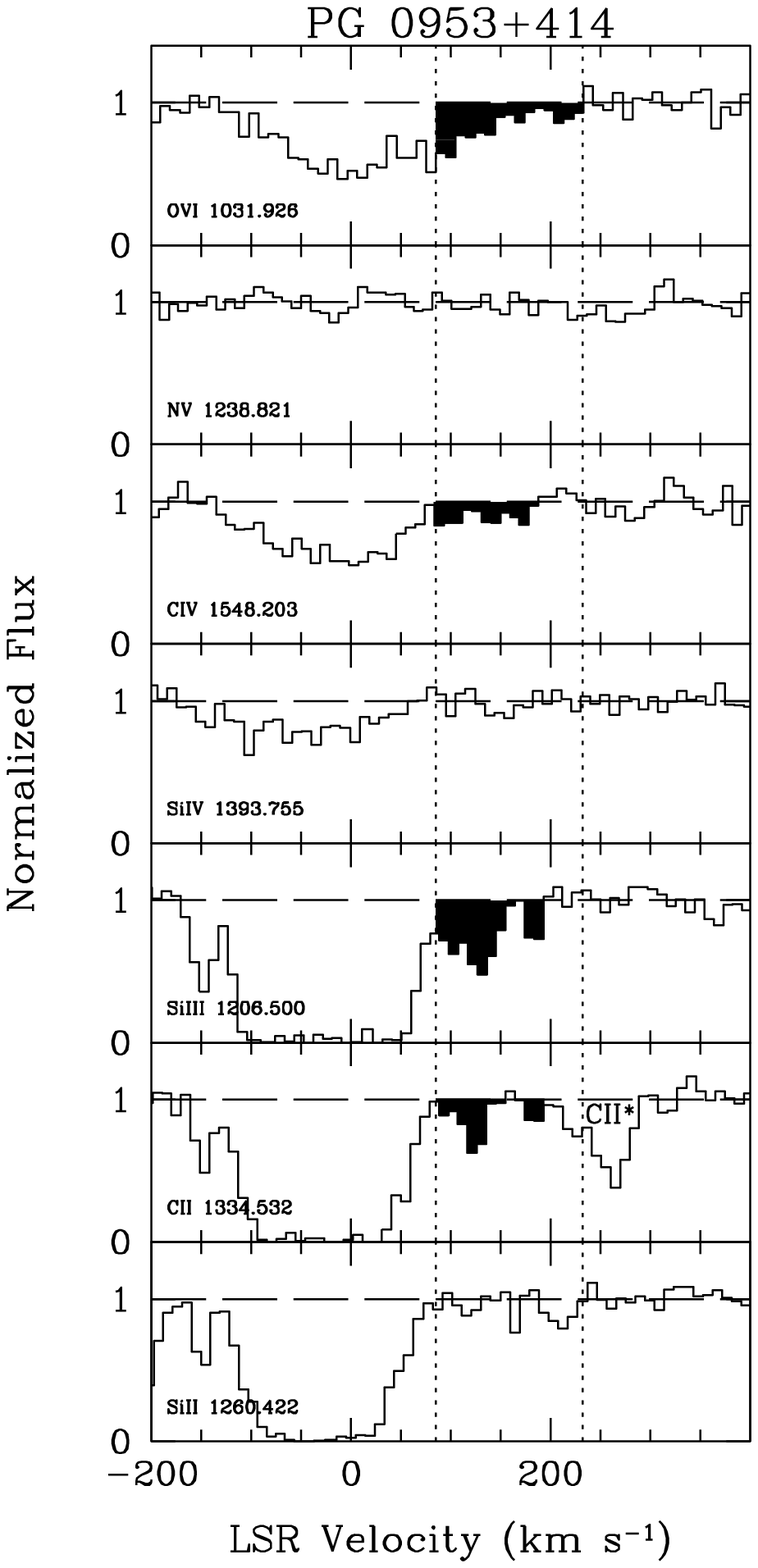}
\caption{Normalized absorption profiles from STIS E140M and {\it FUSE}
data for PG 0953+414.
The shaded regions indicate absorption from the
high-velocity component,
while the vertical dashed lines indicate the velocity extent of the
high-velocity \ion{O}{6} absorption ($85$ to $230$ \kms).
The high-velocity component
can be detected in lines of \ion{O}{6}, \ion{C}{4}, \ion{Si}{3}, and
\ion{C}{2}.  Note the HVC at $V_{\rm LSR}=-150$ km
s$^{-1}$, which is detected in \ion{Si}{3}, \ion{Si}{2}, and
\ion{C}{2}, but not in \ion{H}{1} 21-cm emission.}
\end{figure*}

\begin{deluxetable}{lrccc}
\tablefontsize{\footnotesize}
\tablecolumns{5}
\tablewidth{0pc}
\tablecaption{SUMMARY OF MEASUREMENTS: PG~0953+414 \label{t6}}
\tablehead{
\colhead{} & \colhead{$\lambda$\tablenotemark{a}} & \colhead{} & \colhead{$W_{\lambda
}$} & \colhead{log $N(X)$\tablenotemark{b}} \\
\colhead{Species} & \colhead{(\AA)} & \colhead{} & \colhead{(m\AA)} &
\colhead{($N$ in cm$^{-2}$)}}
\startdata
\ion{O}{6}  & 1031.926 \ \ & &  $79\pm12$ & $13.86^{+0.06}_{-0.06}$ \\
\ion{N}{5}  & 1238.821 \ \ & &   $<34$    & $<13.21$                \\
\ion{C}{4}  & 1548.203 \ \ & &  $62\pm11$ & $13.22^{+0.07}_{-0.08}$ \\
\ion{Si}{4} & 1393.755 \ \ & &   $<29$    & $<12.53$                \\
\ion{Si}{3} & 1206.500 \ \ & & $122\pm11$ & $12.85^{+0.04}_{-0.04}$ \\
\ion{C}{2}  & 1334.532 \ \ & &  $58\pm10$ & $13.52^{+0.06}_{-0.08}$ \\
\ion{Si}{2} & 1260.422 \ \ & &   $<36$    & $<12.38$
\enddata
\tablenotetext{a}{Wavelengths and oscillator strengths are from Morton (2003). }
\tablenotetext{b}{All column densities are calculated
through the apparent optical depth method.  
The HVC occupies the velocity
range $85<V_{LSR}<230$ km s$^{-1}$ based on the various line profiles.
All upper limits are 3$\sigma$ levels.}
\end{deluxetable}

\begin{figure*}
\figurenum{7}
\epsscale{0.58}
\plotone{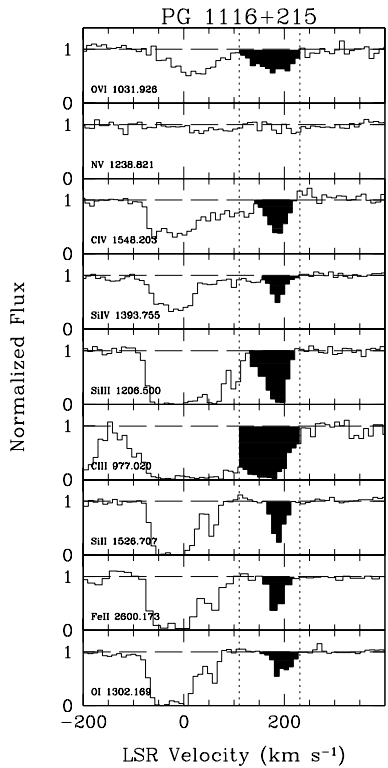}
\caption{Normalized absorption profiles from STIS E140M, STIS E230M,
and {\it FUSE} data for PG 1116+215.  The shaded regions indicate
absorption from the high-velocity component, while the vertical
dashed lines indicate the velocity extent of the high-velocity
\ion{O}{6} absorption ($110$ to $230$ \kms).  Some of the shaded region in
the \ion{O}{1}$\lambda1302.169$ profile is contaminated by Ly$\alpha$ in the
IGM at $z=0.07118$.  Including the profiles
shown here, the high-velocity component is detected in \ion{O}{6},
\ion{C}{4}, \ion{Si}{4}, \ion{C}{3}, \ion{Si}{3}, \ion{N}{3}, 
\ion{C}{2}, \ion{Si}{2},
\ion{Fe}{2}, \ion{N}{2}, \ion{Mg}{2}, and \ion{O}{1}.}
\end{figure*}

High-quality \FUSE and STIS E140M data exist for the PG 0953+414
sight line.  Fox et al. (2005, in preparation) presents a very detailed
study of high-velocity gas in this sight line.  
Profiles of absorption lines for several ion species are
shown in Figure 6.  The \ion{O}{6} profile shows positive high-velocity
absorption extending to a maximum velocity of
$V_{\rm LSR}=230$ \kms.  Based on its \ion{O}{6} profile, S03
identify the high-velocity component as a positive-velocity wing to
Galactic absorption, possibly tracing hot Galactic fountain gas.
Detections of lower ions in the STIS bandpass indicate that this
identification is not likely and suggest that the high-velocity absorption
traces a discrete HVC.  Absorption is clearly detected in \ion{C}{4} and 
\ion{Si}{3}, extending over a more narrow velocity range than the
high-velocity \ion{O}{6} absorption.  We use the minimum velocity
cutoff for the \ion{C}{4} and \ion{Si}{3} absorption to establish
\ion{O}{6} integration limits of $85<V_{\rm LSR}<230$ \kms.
High-velocity absorption is detected in \ion{C}{2} $\lambda1334.53$ as
well, although some contamination from \ion{C}{2}$^{*}\lambda1335.71$ is
likely at $V_{\rm LSR}>200$ \kms.  Since the absorption features
up to $V_{\rm LSR}=195$ \kms
are also detected in \ion{Si}{3} and \ion{C}{4}, we can attribute absorption
in this velocity range to high-velocity \ion{C}{2} with high confidence.
We do not believe the absorption centered at $V_{\rm LSR}=210$ \kms
in the \ion{Si}{2} $\lambda1260.42$ profile is from high-velocity gas,
since the feature is not detected in other profiles.  Because of
blending from saturated Galactic absorption, this HVC has not been studied
in \ion{C}{3} $\lambda977.02$.
Column density measurements for the HVC detected in \ion{O}{6} over
the range $85<V_{\rm LSR}<230$ \kms are shown in Table 6.

Finally, we note the new detection of a negative-velocity HVC in this
sight line centered at $V_{\rm LSR}=-150$ \kms, possibly from
an extended atmosphere of HVC Complex M.  Although the HVC cannot be
detected in \ion{H}{1} 21-cm emission (Wakker \etal 2003), it is clearly
detected in \ion{Si}{2}, \ion{C}{2}, and \ion{Si}{3}.  Interestingly, this
HVC is not detected in \ion{O}{6}.

\placetable{t6}

\subsection{PG~1116+215}

\begin{figure*}
\figurenum{8}
\epsscale{0.6}
\plotone{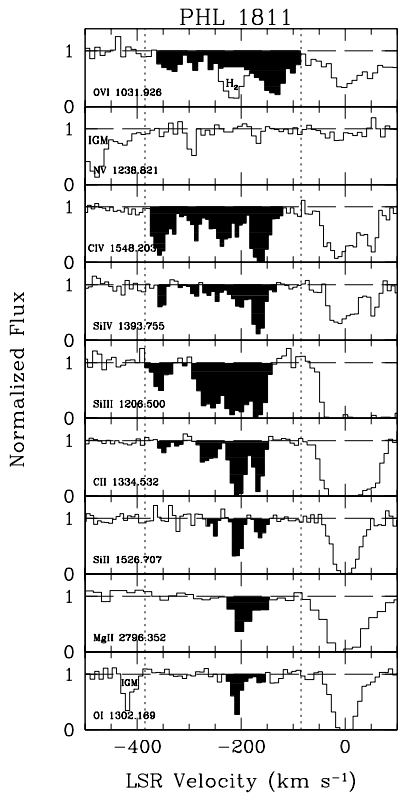}
\caption{Normalized absorption profiles from STIS E140M, STIS G230MB,
and {\it FUSE} data for PHL 1811.
The shaded regions indicate
absorption from high-velocity components, while the vertical
dashed lines indicate the velocity extent of the high-velocity
absorption from the various resonance lines ($-385$ to $-85$ \kms).
High-velocity absorption is detected in \ion{O}{6}, \ion{C}{4}, \ion{Si}{4},
\ion{Si}{3}, \ion{C}{2}, \ion{Si}{2}, \ion{Mg}{2}, and \ion{O}{1}.  }
\end{figure*}

High-quality data for PG~1116+215 from both \FUSE and \HST are now
publicly available.  The IGM absorbers have been analyzed by Sembach
\etal (2004).  The \HST STIS observations include data using the E140M
grating, as well as the E230M grating covering a wavelength range
$2010-2820$ \AA.  The \ion{O}{6} profile indicates a high-velocity
component spaning the velocity range $110-230$ \kms.  In
addition to \ion{O}{6}, high-velocity absorption from this component
can be detected in at least 22 resonance lines of \ion{C}{4},
\ion{Si}{4}, \ion{C}{3}, \ion{Si}{3}, \ion{N}{3}, \ion{C}{2}, \ion{Si}{2},
\ion{Fe}{2}, \ion{N}{2}, \ion{Mg}{2}, and \ion{O}{1}.  Profiles for
several key absorption lines are shown in Figure 7.  We include this
sight line in our sample only to present the fact that this HVC can be
detected in lines of various ionization stages.  If the HVC is
characterized by a metallicity of $Z/Z_{\sun}=0.1$ in an abundance
pattern of Grevesse \& Sauval (1998), then the optical depth of the
high-velocity \ion{O}{1} $\lambda1302.17$ feature suggests a column
density $N$(\ion{H}{1}) $\sim 10^{18}$ cm$^{-2}$.  Such a value of
$N$(\ion{H}{1}) is barely below the detection limit for single-dish
radio observations.  Ganguly \etal (2005) report that some
of the \ion{O}{1} profile is contaminated by Ly$\alpha$ at $z=0.07118$, 
although
they do confirm that at least half of the absorption feature's optical depth 
arises from high-velocity \ion{O}{1}.  The \ion{O}{1} $\lambda1039.23$
would be useful for confirming the detection, but the column density is too
low to detect the HVC in this line.

The \ion{H}{1} can be measured for the HVC in saturated Lyman lines
from Ly$\eta$ to Ly$\lambda$ ($n = 12 \rightarrow 1$), where
absorption from the HVC begins to deblend from Galactic absorption.
The measured $N$(\ion{H}{1}) from a curve-of-growth fit is poorly 
constrained; using the observed
equivalent widths of the Lyman lines, 
we find log~$N$(\ion{H}{1})$ = 16.5-18.5$, based on
$1\sigma$ error bars.
Ganguly \etal (2005) find a range from 16.72 (from direct integration
of Ly$\lambda$ at 918.13~\AA) to 18.3 (from the Wakker \etal [2003]
non-detection of 21 cm emission).  Their best-fit value is
log~$N$(\ion{H}{1}) = 17.82, with a metallicity
[O/H] $= -0.69^{+0.39}_{-0.16}$ from O~I $\lambda1302$.

\subsection{PHL~1811}

PHL~1811 has good quality \FUSE data (Jenkins \etal 2003), as well as
\HST STIS E140M and G230MB data.  At the time of our analysis, some of the
STIS E140M data was proprietary (program ID 9418). The data we present
includes only
half of the total exposure in the archive.  Data for several important
resonance lines are shown in Figure 8.  The \ion{O}{6} profile shows a
broad, high-velocity component with a width $\sim300$ \kms.
Contamination from interstellar H$_{2}$ L(6-0)P(3) is present,
although some of the underlying absorption arises from high-velocity
\ion{O}{6}.  In lower ionization species, the high-velocity component
is detected in \ion{C}{4}, \ion{Si}{4}, \ion{Si}{3}, \ion{C}{2},
\ion{Si}{2}, \ion{Mg}{2}, and \ion{O}{1}.  Several of the profiles
indicate velocity structure for the HVC, with evidence for as
many as four high-velocity sub-components.  This sightline analysis
will benefit from the additional exposures in the \HST archive,
to better study the complicated sub-component absorption.
We defer a thorough analysis of this
sight line until the full dataset is available.  We include the sight
line in our sample to present the positive detection of low ion
absorption associated with the \ion{O}{6} HVC.  The \ion{O}{1} detection
suggests that the HVC has a column density of
$N$(\ion{H}{1}) $\sim10^{18}$ cm$^{-2}$.

\subsection{PKS~1302$-$102}

\begin{figure*}
\figurenum{9}
\epsscale{0.7}
\plotone{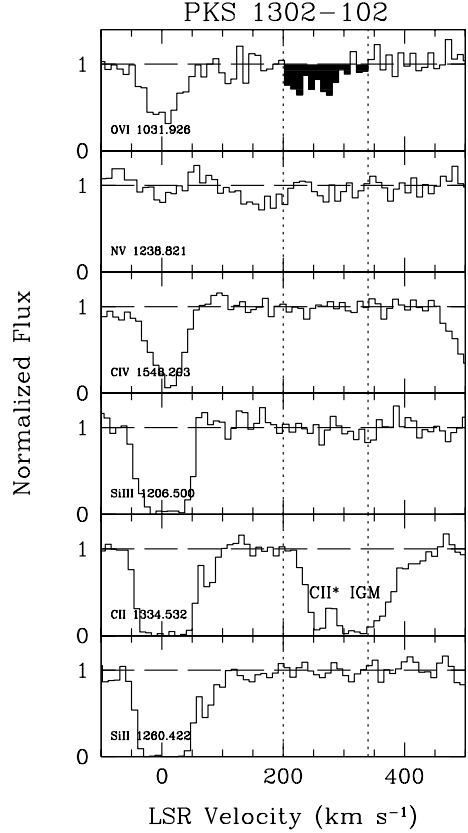}
\caption{Normalized absorption profiles from STIS E140M and and {\it
FUSE} data for PKS 1302$-$102.  The vertical dashed lines indicate the
velocity extent of the high-velocity component determined from its
\ion{O}{6} absorption ($200$ to $340$ \kms).  The shaded region
indicates absorption from the high-velocity component, which in
this case can only be detected for the \ion{O}{6} line.  }
\end{figure*}

\begin{deluxetable}{lrccc}
\tablefontsize{\footnotesize}
\tablecolumns{5}
\tablewidth{0pc}
\tablecaption{SUMMARY OF MEASUREMENTS: PKS~1302$-$102 \label{t7}}
\tablehead{
\colhead{} & \colhead{$\lambda$\tablenotemark{a}} & \colhead{} & \colhead{$W_{\lambda
}$} & \colhead{log $N(X)$\tablenotemark{b}} \\
\colhead{Species} & \colhead{(\AA)} & \colhead{} & \colhead{(m\AA)} &
\colhead{($N$ in cm$^{-2}$)}}
\startdata
\ion{O}{6}  & 1031.926 \ \ & &  $87\pm21$ & $13.91^{+0.09}_{-0.11}$ \\
\ion{N}{5}  & 1238.821 \ \ & &   $<47$    & $<13.35$         \\
\ion{C}{4}  & 1548.203 \ \ & &   $<40$    & $<13.00$                \\
\ion{Si}{3} & 1206.500 \ \ & &   $<56$    & $<12.42$                \\
\ion{Si}{2} & 1260.422 \ \ & &   $<46$    & $<12.45$
\enddata
\tablenotetext{a}{Wavelengths and oscillator strengths are from Morton
   (2003). }
\tablenotetext{b}{All column densities are calculated
through the apparent optical depth method.  
The HVC occupies the velocity
range $200<V_{LSR}<340$ km s$^{-1}$ based on the \ion{O}{6} profile.
All upper limits are 3$\sigma$ levels.}
\end{deluxetable}

The \FUSE and \HST STIS data for the PKS 1302$-$102 sight line are of
high quality, with line profiles for several absorption lines shown in
Figure 9.  Absorption is detected in the \ion{O}{6} profile at $1032.2$~\AA,
which would cover the velocity range $200<V_{\rm LSR}<340$ km~s$^{-1}$.
Absorption in the \ion{C}{2} $\lambda1334.53$ profile in the HVC velocity
range is contaminated by IGM absorption.  The feature centered near
$V_{\rm LSR} =260$ \kms arises from \ion{C}{2}*$\lambda1335.71$,
while the broader feature at higher velocity shows little correlation with
the \ion{O}{6} profile and is intergalactic Ly$\alpha$ at $z = 0.099$, 
confirmed by the detection of Ly$\beta$ at 1127.2~\AA.  The 
feature at $V_{\rm LSR}=160$ \kms in the \ion{N}{5} profile is not likely
an HVC, since it is considerably stronger than the Galactic absorption 
and cannot be detected in the weaker line of the doublet, 
\ion{N}{5} $\lambda1242.80$.  The upper limit is
N(N~V)/N(O~VI) $< 0.3$. Corresponding absorption cannot
be detected in lower ions, raising the possibility that the
high-velocity feature may arise from IGM absorption in a resonance
line other than the \ion{O}{6}. There are several low-redshift
Ly$\alpha$ features detected in the STIS spectrum, which we use to
assess this possibility.  We find that, for the redshifts of the
observed Ly$\alpha$ absorbers, the data are inconsistent with the
feature at $1032.2$ \AA\ arising from other resonance lines at the
those redshifts.  We therefore confirm that this is a high-velocity
\ion{O}{6} detection. {\it PKS~1302$-$102 and 3C~273 are the only cases
where a discrete high-velocity \ion{O}{6} absorber cannot be detected
in other ion species.}  Column density measurements and upper limits
for this HVC are shown in Table 7.

\placetable{t7}

\subsection{Ton~S180}

\begin{figure*}
\figurenum{10}
\plotone{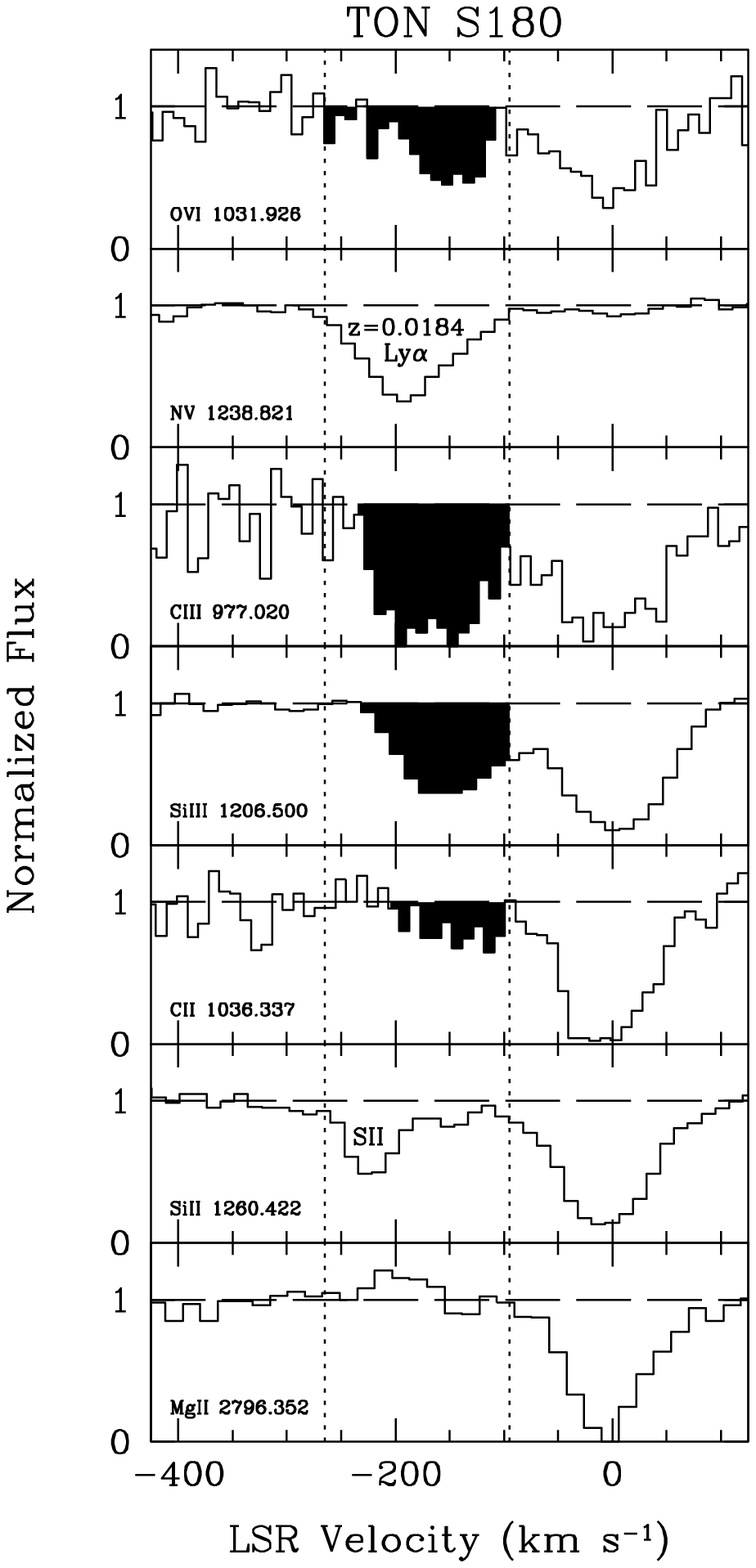}
\caption{Normalized absorption profiles from STIS G140M and \FUSE data
for Ton~S180.  The shaded regions indicate
absorption from the high-velocity component, while the vertical
dashed lines indicate the velocity extent of the high-velocity
\ion{O}{6} absorption ($-265$ to $-95$ \kms).
The high-velocity component can be detected in lines of \ion{O}{6},
\ion{C}{3}, \ion{Si}{3}, and \ion{C}{2}. We have no \ion{C}{4} or \ion{Si}{4}
coverage for
this sight line.}
\end{figure*}

\begin{deluxetable}{lrccc}
\tablefontsize{\footnotesize}
\tablecolumns{5}
\tablewidth{0pc}
\tablecaption{SUMMARY OF MEASUREMENTS: TON~S180 \label{t8}}
\tablehead{
\colhead{} & \colhead{$\lambda$\tablenotemark{a}} & \colhead{} & \colhead{$W_{\lambda
}$} & \colhead{log $N(X)$\tablenotemark{b}} \\
\colhead{Species} & \colhead{(\AA)} & \colhead{} & \colhead{(m\AA)} &
\colhead{($N$ in cm$^{-2}$)}}
\startdata
\ion{O}{6}  & 1031.926 \ \ & & $162\pm26$ & $14.26^{+0.06}_{-0.06}$ \\
\ion{C}{3}  &  977.020 \ \ & & $316\pm34$ & $>14.09$                \\
\ion{Si}{3} & 1206.500 \ \ & & $252\pm8$  & $13.23^{+0.01}_{-0.01}$ \\
\ion{C}{2}  & 1036.337 \ \ & &  $76\pm17$ & $13.91^{+0.08}_{-0.11}$  \\
\ion{Mg}{2} & 2796.352 \ \ & &   $<173$   & $<12.60$
\enddata
\tablenotetext{a}{Wavelengths and oscillator strengths are from Morton (2003).}
\tablenotetext{b}{All column densities are calculated
through the apparent optical depth method.  
The HVC occupies the velocity
range $-265<V_{LSR}<-95$ km s$^{-1}$ based on the various line profiles.
All upper limits are 3$\sigma$ levels.}
\end{deluxetable}

The Ton~S180 sight line has good quality \FUSE and \HST STIS G140M data.
STIS G230MB observations also exist, although the low quality of that data
allows only upper limits to be established on the column density of \ion{Mg}{2}.
The raised continuum near $V_{\rm LSR} = -200$ \kms in the \ion{Mg}{2} profile
indicates that contamination may be present, although it does not appear
sufficiently significant to dominate any possible $>3\sigma$ high-velocity
absorption, which should be cenetered near $V_{\rm LSR}\approx-150$ \kms.
Line profiles for several ion species are shown in Figure 10.  We detect
high-velocity absorption in
\ion{O}{6}, \ion{C}{3}, \ion{Si}{3}, and \ion{C}{2}.  Some of the
absorption near $V_{\rm LSR} = -215$ \kms in the \ion{O}{6} profile
could arise from \Htwo L(6-0)P(3), although an analysis of other
\Htwo $J=3$ lines indicates that the contribution should be negligible.

The O~VI high-velocity component appears well separated from the Galactic 
component, although blending is apparent in the
\ion{Si}{3} profile.  Based on the various absorption profiles, we
adopt an integration range of $-265 < V_{\rm LSR} <-95$ \kms  for the
high-velocity component.  Column density measurements for the HVC are presented
in Table 8.  There is also O~VI absorption at 200--280 \kms (S03), which we do
not measure because it may be associated with three galaxies in the
Coma-Sculptor Group (Wakker \etal 2003).
The strong feature in the \ion{N}{5} profile, which occupies a
nearly identical velocity range as the high-velocity component, is actually
IGM Ly$\alpha$ absorption at $z=0.0184$ (Penton \etal 2004).  This
identification is confirmed by a corresponding Ly$\beta$ detection at
this redshift near 1044.6 \AA.  Since saturation is clearly present in the
\ion{C}{3} profile, we can establish only a lower limit on $N$(\ion{C}{3}).
The high-velocity range of the \ion{Si}{2} $\lambda1260.42$ profile is
contaminated by Galactic \ion{S}{2} $\lambda1259.52$.  Blending makes
it impossible to accurately measure $N$(\ion{Si}{2}) for the HVC, although
the feature near $V_{\rm LSR} = -140$ \kms is probably high-velocity
\ion{Si}{2}.  With the current data, we do not have access to strong
Si~II lines (e.g., 1190.416, 1193.290, 1526.707). 

\placetable{t8}

\subsection{UGC~12163}

\begin{figure*}
\figurenum{11}
\epsscale{0.67}
\plotone{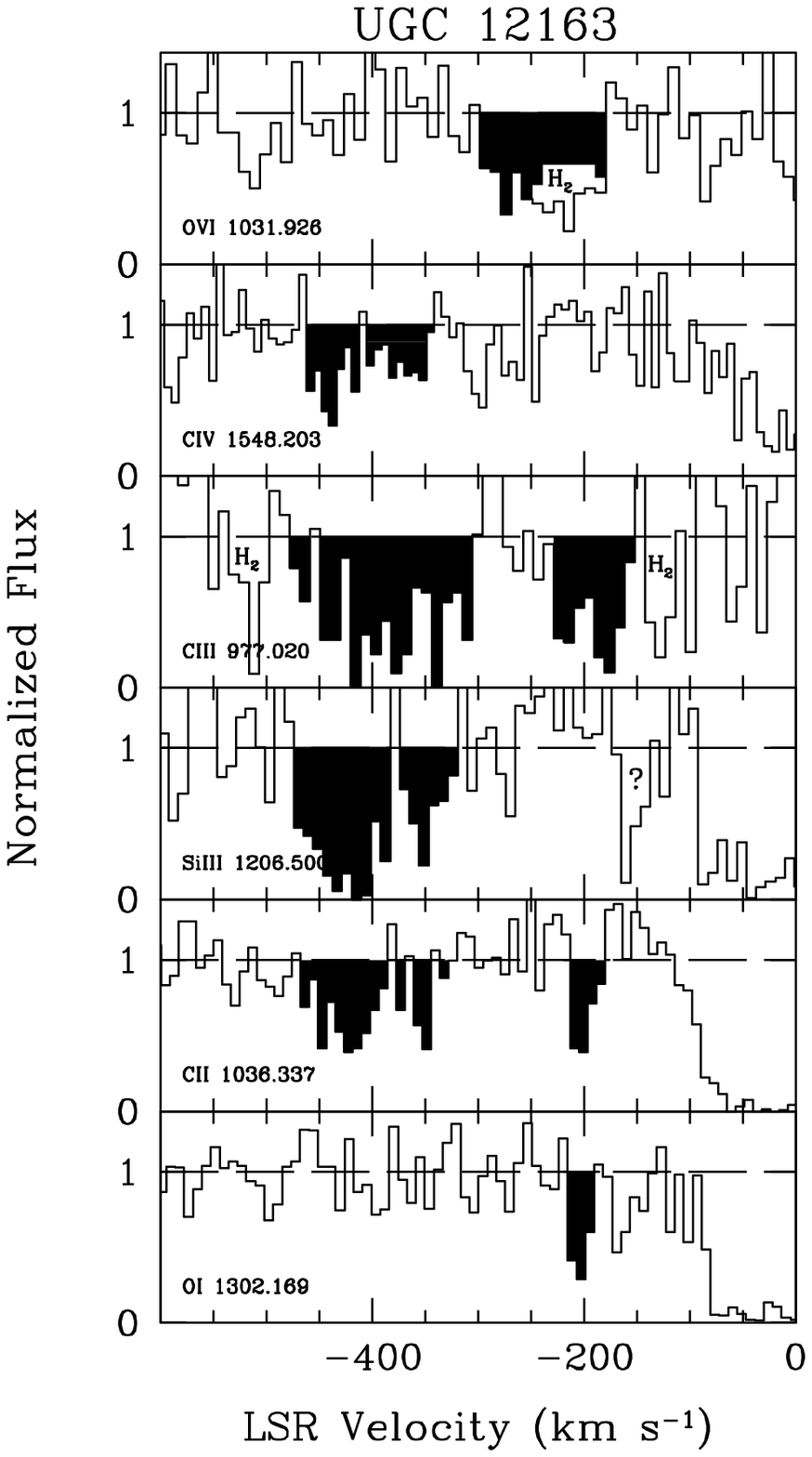}
\caption{Normalized absorption profiles from STIS E140M and
\FUSE data for UGC~12163.  Shaded regions indicate possible
high-velocity absorption,
which, in this case, may be detected in \ion{O}{6}, \ion{C}{4}, \ion{C}{3},
\ion{Si}{3}, \ion{C}{2}, and \ion{O}{1}.  Owing to poor data quality and
complicated absorption features, we have
made no attempt to determine integration ranges and column densities
for the high-velocity absorption.}
\end{figure*}

Although \FUSE and \HST STIS data exist for UGC 12163, the low flux of this
target produces the lowest S/N data in our sample.
In spite of the poor data quality, high-velocity absorption can be
detected in several ion species, ranging in ionization state from
\ion{O}{1} to \ion{O}{6}, including \ion{C}{2}, \ion{C}{3}, and \ion{C}{4}.
Profiles of resonance lines where high-velocity absorption can be detected
are shown in Figure 11.  We have subtracted off contamination from \Htwo
L(6-0)P(3), despite the poor data quality.  Although the resulting profile
indicates an imprecise subtraction, it is a useful exercise in
demonstrating that some of the underlying absorption must be from \ion{O}{6}.
After H$_{2}$ subtraction, the high-velocity \ion{O}{6} is detected primarily
over the velocity range $-300 < V_{\rm LSR} < -180$ \kms, while the lower
ionization species show a more complicated high-velocity structure.
We detect with certainty absorption in \ion{C}{3}, \ion{C}{2}, and \ion{O}{1}
at similar velocity as the \ion{O}{6}, although over a much narrower
velocity width.  Additionally, we detect high-velocity \ion{C}{4},
\ion{Si}{3}, \ion{C}{3}, and \ion{C}{2} over a velocity range
$-480<V_{\rm LSR}<-300$ \kms, with no corresponding \ion{O}{6}
absorption.  This complicated absorption structure underscores the
need for better quality \FUSE and \HST data for this sight
line since many features are difficult to distinguish from noise.
Because of the noisy data and the difficulty in establishing the velocity
range of the absorption features, we do not attempt to measure column
densities.  The data are included solely to present the
detection of lower ionization species associated with the \ion{O}{6}
HVC.

\section{DISCUSSION}

\subsection{Kinematics and Spatial Distribution}

Given the similar ionization characteristics of most HVCs in our 
sample, it is instructive to consider the spatial distribution of the
sight lines.  In Figure 1, we showed the locations of the sight lines
in our sample on an all-sky map of the highly ionized HVCs from the
S03 survey.  Also labeled are the Local Group barycenter at
$(\ell, b) = (147\arcdeg,-25\arcdeg)$ (Karachentsev \& Makarov 1996) and
the direction of the Milky Way's motion, $(\ell, b) = (107\arcdeg,-18\arcdeg)$
at $V=90$ \kms (Einasto \& Lynden-Bell 1982) with respect to the
Local Group barycenter. The HVCs in the figure are classified as either
``Local Group'' (PKS 2155$-$304, Mrk 509, MRC 2251$-$178, PHL 1811, Ton S180), 
``extreme positive (north)'' (PG 0953+414, PG 1116+215, 3C 273, 
PKS 1302$-$102), or ``Magellanic Stream
extension'' (Mrk 1513, UGC 12163, HE 0226$-$4110) 
in S03's identification system.  None of the HVCs shown are
detected in \ion{H}{1} emission.  The 12 sight lines discussed in this
paper do not cluster in a restricted portion of the sky and probably 
represent the spatial distribution of the full population.

The segregation of the positive- and negative-velocity highly ionized HVCs
and the apparent symmetry of this pattern with regard to the location of 
the Local Group barycenter led Nicastro \etal (2003) to propose that the
\ion{O}{6} HVCs trace the local WHIM filament.  The kinematics of the
population could be explained if the Galaxy is drifting through 
a hot gaseous medium in the Local Group, although there are other models that
are consistent with the observations.  For example, an infalling shell of 
gas, spherically symmetric about the Galactic center, produces a similar 
velocity segregation in the LSR.  This model is not unique, and it could 
be refined by including additional features:   
(1) the range of O~VI velocities rather than just the centroid;
(2) asymmetric infall velocities and shell radii; and  
(3) peculiar motions from SNRs and the Galactic fountain.  
Nevertheless, the model illustrates the basic features 
of gaseous infall encountering the enhanced pressure of a rotating halo, 
without including details of shocks and deceleration of the flow.

As a simple illustration, we examined the all-sky velocity distribution 
of a toy model of a radially infalling shell at Galactocentric
radius, $R=15$ kpc, $V_{\rm infall}=50$ \kms, and assuming standard 
Galactic rotation, $V_{\rm rot}=220$ \kms towards $\ell = 90\arcdeg$
at $R_{\sun}=8.5$ kpc.  Such a model can explain the general trend of
positive velocities at $180\arcdeg< \ell <360\arcdeg$ and negative
velocities at $0\arcdeg< \ell <180\arcdeg$, although an anomaly is
the clustering of positive velocity HVCs near $\ell \approx180\arcdeg$, 
the direction of the Galactic anti-center.  Because
the LSR velocities of clouds in this direction are unaffected by Galactic
rotation, they should measure velocities with respect to the
Galactic disk.  However, the clouds near $\ell = 180\arcdeg$ are clearly 
not infalling.  A recent analysis (Savage \etal 2005) of a highly ionized 
HVC toward Mrk 421, at $(\ell, b) = (179.8\arcdeg,65.03\arcdeg)$ concludes 
that the HVC most likely traces outflowing gas in a Galactic
fountain or H~I supershell (Heiles 1979).  The redshifted clouds near
$\ell \approx 180\arcdeg$ (Fig.\ 1) are at high Galactic latitude, so
an association with such features would not be surprising.  There is
also the more exotic possibility that some of this gas is falling
toward the Local Group barycenter.

The locations of the sight lines and kinematics of the HVCs can also
be compared to the distribution of known high-velocity objects in
all-sky \ion{H}{1} maps (e.g., Figure 16 from Wakker \etal 2003 and 
Figure 11a from S03).  The negative-velocity HVCs in our sample are
clustered in the southern Galactic hemisphere between
$17\arcdeg< \ell <140\arcdeg$.  All the negative-velocity HVCs in the
sample except Mrk 1513 (detected in \ion{C}{3} and \ion{Si}{3}) are
detected in singly-ionized species.  These sight lines lie within
$\sim20\arcdeg$ of either the \ion{H}{1}-detected portions of the
Magellanic Stream\footnotemark\footnotetext{The Magellanic Stream has 
an average angular width of $15\arcdeg$, running from $(\ell,b) = 
(285\arcdeg, -30\arcdeg)$, through the south Galactic pole, to 
$(\ell,b) = (90\arcdeg, -40\arcdeg)$.} or the Galactic Center 
Negative (GCN)\footnotemark\footnotetext{The GCN clouds
are a collection of negative-velocity HVCs bounded by 
$20\arcdeg < \ell < 45\arcdeg$ and $-40\arcdeg < b < -10\arcdeg$.} 
clouds and they share similar kinematics to those objects.  These
negative-velocity, highly ionized HVCs may trace extended low column
density ``atmospheres" of the Magellanic Stream or GCN clouds,
produced by photoionization or interfaces arising from thermal
conduction or turbulent mixing (Indebetouw \& Shull 2004). Evidence 
from studies of high and low ions in Complex C (Fox et al. 2004) as
well as the correlation between OVI and mapped HI HVCs
(S03) indicate that halo HVCs have hot, highly ionized outer 
skins or interfaces, although the extent of these atmospheres is unknown.  
These blueshifted highly ionized HVCs are {\it not} associated with the 
Anti-Center Complex at $-210$ \kms seen in 21-cm emission in the southern
Galactic hemisphere (Mirabel 1982).

The positive-velocity, highly ionized HVCs in the northern Galactic
hemisphere present a more difficult problem, since few of these
objects are kinematically related to \ion{H}{1} structures in their
regions of sky.  Only the positive-velocity HVC toward HE 0226$-$4110,
which lies within $\sim15\arcdeg$ of the Magellanic Stream, shares a
kinematic connection to an extension of the Magellanic Stream.  As
stated previously, several of these clouds do not conform to a simple
infall model and may represent Galactic fountain material.  The 3C 273
\ion{O}{6} wing is consistent with a Galactic fountain (Sembach \etal
2001), since the sightline passes near Galactic Radio Loops I and IV of
the North Polar Spur (Burks \etal 1994).  The PKS 1302$-$102 HVC is
also detected only in \ion{O}{6} and lies $\sim20\arcdeg$ from 3C 273.
The large positive velocity of this HVC (200 -- 340 \kms) and the
lack of a connection to Galactic absorption suggest that a fountain
origin is unlikely.  The HVCs near $\ell = 180\arcdeg$ in the
PG~0953+414 and PG~1116+215 sight lines have significant
low-ion detections associated with the \ion{O}{6}.  It is unlikely
that these clouds trace a Galactic fountain, unless large amounts of
singly ionized and neutral gas are entrained in the fountain and are
somehow able to maintain temperatures as low as $T\sim10^{4}$ K.  The low
metallicity of this HVC, [O/H] = $-0.69^{+0.39}_{-0.16}$ (Ganguly
\etal 2005), suggests that such a scenario is unlikely, as the
metallicity of fountain gas is likely to be near solar.  While the
origin of the positive-velocity, highly ionized HVCs is unknown, 
we note that a simple infall model can reproduce some of the general 
kinematics of the full population.  We are undertaking more detailed
modeling of the range of O~VI velocities, and allowing some of the
model parameters to vary.

\subsection{Ion Detections in Highly Ionized HVCs}

\begin{deluxetable}{lccc}
\tablefontsize{\footnotesize}
\tablecolumns{4}
\tablewidth{0pc}
\tablecaption{Absorption Detections for O~VI HVCs\label{t9}}
\tablehead{
\colhead{Sightline} & \colhead{[C,Si]~IV} & \colhead{[C, Si]~III} & \colhead{[C, Si]~II} }
\startdata
3C 273  & & & \\
HE 0226$-$4110 & $\times$ & $\times$ & $\times$ \\
MRC 2251-178 & n/c\tablenotemark{b} & $\times$ & $\times$ \\
Mrk 509 \tablenotemark{a} & $\times$ & $\times$ & $\times$ \\
Mrk 1513 & n/c & $\times$ &  \\
PG 0953+414 & $\times$ & $\times$ & $\times$ \\
PG 1116+215 & $\times$ & $\times$ & $\times$ \\
PHL 1811 & $\times$ & $\times$ & $\times$ \\
PKS 1302-102 & & & \\
PKS 2155-304\tablenotemark{a} & $\times$ & $\times$ & $\times$ \\
Ton S180 & n/c & $\times$ & $\times$ \\
UGC 12163 & $\times$ & $\times$ & $\times$
\enddata
\tablenotetext{a}{From Collins et al. (2004)}
\tablenotetext{b}{``n/c'' denotes no spectral coverage for the sight lines
observed only with the HST-STIS G140M grating.}
\end{deluxetable}

Including the two sight lines from CSG04, we have now surveyed highly ionized
HVCs in 12 QSO sight lines.  These sight lines, selected from the detection
of high-velocity \ion{O}{6} and the availability of \HST STIS data, are
investigated primarily through detections of single, double, and
triple ions of C and Si.  Some of these HVCs are detected in
\ion{O}{1} as well.  We made no definite detection of N~V, a weak ion whose
measurement is almost always difficult (Indebetouw \& Shull 2004).
In Table 9 we summarize the ion detections for the HVCs in these sight lines.
In the 12 sight lines, we detect low ions (\ion{C}{2} or \ion{Si}{2}) in 9 
(75\%) of the cases, and either C~III or Si~III in 10 of 12 sight lines.
In one of the three cases where low ions cannot be detected, the HVC toward
Mrk 1513, absorption is detected in doubly-ionized species;  \ion{Si}{3} and
\ion{C}{3} are detected in the Mrk 1513 HVC, while the 3$\sigma$
detection limit for \ion{C}{2} $\lambda 1036.34$ is well above all but one
measured $N$(\ion{C}{2}) for the HVCs in this sample.  In better quality data,
it is likely that the absorption characteristics for the Mrk~1513 HVC would
be similar to the other HVCs detected in low ions.

Two HVCs in the sample, toward 3C~273 and PKS~1302$-$102, are not detected in
any other ion besides \ion{O}{6}.  The high-velocity \ion{O}{6} 
towards 3C~273 is a positive-velocity wing to Galactic absorption, and can be
explained as cooling gas from a Galactic fountain (Sembach \etal 2001) or
SNR radio shells (Burks \etal 1994).  The PKS 1302$-$102 high-velocity
\ion{O}{6}\ absorption is well separated from the Galactic component; one
possible interpretation is that it traces a low-$z$ filament of the WHIM.
X-ray observations of O~VII or O~VIII in this sight line
would be useful in investigating this possibility, although the source is
probably too faint for current X-ray spectrographs.

\placetable{t9}

\subsection{Modeling the Column Density Pattern}

\begin{figure*}
\figurenum{12}
\epsscale{1}
\plotone{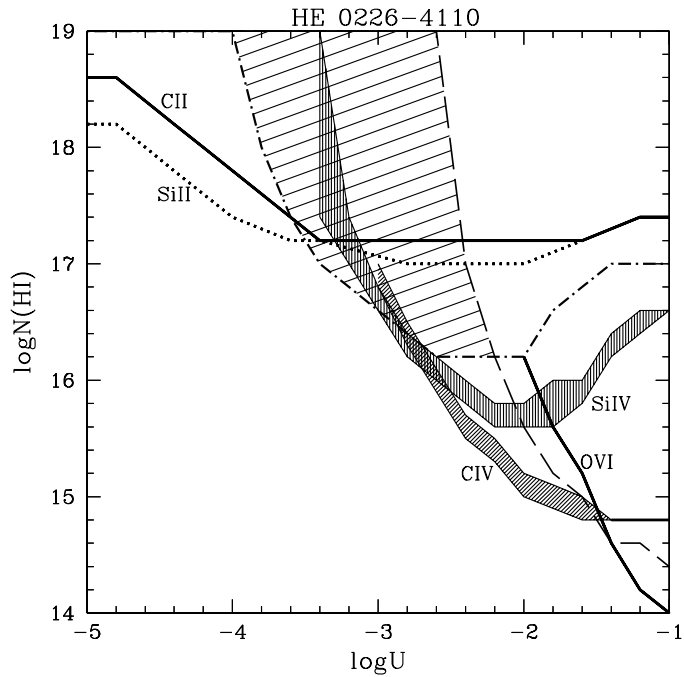}
\caption{Grid of CLOUDY model calculations with model constraints for
the HVC toward HE~0226$-$4110.  The photoionization model is characterized by
a QSO power-law spectrum with spectral index, $\alpha=1.8$, an intensity
log~$\phi=4.0$ (photons cm$^{-2}$ s$^{-1}$), and a metallicity $Z=0.1 Z_{\sun}$.
For the constaints, we use the $2\sigma$ range of the measured ion column density,
the measured value as a lower limit if saturation is present, or the upper limit.
The constraints on the models based on detections of \ion{C}{2}, \ion{Si}{2},
\ion{Si}{4}, and \ion{O}{6} are represented by the labeled curves.  The large
shaded region in the upper center part of the figure is the constraint based on
the lower limit to $N$(\ion{Si}{3}) (dot-dashed line) and the upper limit to
$N$(\ion{N}{5}) (dashed line).  The \ion{C}{2} and \ion{Si}{2} measurements
are consistent with photoionization in an HVC with log~$N$(\ion{H}{1})$=17.2$
and log~$U=-3.4$ (log~$n_{H}=-3.1$).}
\end{figure*}

As demonstrated in the previous section, most of the HVCs discussed in this
work exhibit a similar ionization pattern to the HVCs studied by CSG04 in
the PKS~2155$-$304 and Mrk~509 sight lines.  A key result from modeling those
HVCs was that the column density pattern cannot be replicated by a
single-phase photoionization model. Although low-ionization species,
defined here as C~II/III/IV and Si~II/III/IV, can be produced simultaneously 
by such a model, the observed columns of \ion{O}{6} are several orders
of magnitude larger than predicted.  By assuming that the low
ions are photoionized, the models place a lower limit on the hydrogen density,
$n_H > 10^{-4}$ cm$^{-3}$, some 20--100 times larger than the predicted 
WHIM gas density, $n_H \approx (1-5) \times 10^{-6}$ cm$^{-3}$.
This WHIM has an overdensity $\delta = 5-10$ (Cen \& Ostriker 1999; 
Dav\'e \etal 1999) relative to the mean density of the low redshift universe,
$\langle n_H \rangle = (1.90 \times 10^{-7}~{\rm cm}^{-3})(1+z)^3$.  
  
At $10^6$~K, the cooling time for hot, low-density WHIM gas at 10\% solar 
metallicity is over 100~Gyr (Shull 2003; CSG04), much longer than
the Hubble time. Thus, it is implausible to suggest that the observed low 
ions arise in cooled portions of WHIM filaments. 
The cooling estimates assume approximate pressure equilibrium throughout the 
clouds.  Since the sound crossing time is $\sim10^8$~yr across 1 kpc,
constant pressure is only an approximation.

Here, we take a similar approach as CSG04 and attempt to model the observed
ionization pattern in this sample of HVCs as arising from an incident
ionizing radiation field.  We can immediately rule out a stellar spectrum,
since such fields are too soft to produce significant fractions of ions higher
than doubly ionized.  The most successful CSG04 model used the
integrated radiation from background QSOs and AGNs.  Such ionizing fields
best reproduce the observed column density ratios in the highly ionized HVCs,
where typically $N$(\ion{C}{2})$\sim N$(\ion{C}{3})$\sim N$(\ion{C}{4}).
As in CSG04, we generated a grid of photoionization models using the code 
CLOUDY (Ferland 1996), making the simplifying assumption that the absorbing
gas can be treated as plane-parallel slabs illuminated by incident
AGN background radiation.  There are three basic free
parameters that go into these models: \ion{H}{1} column density, gas
number density, and gas metallicity.  We assume the gas metallicity to
be 10\% solar, similar to that observed (CSG03) in other HVCs
($Z = 0.1-0.4~ Z_{\sun}$), with a relative abundance pattern assumed to be 
solar.  In CSG04, we worked with a smaller parameter space, since the 
\ion{H}{1} column densities in the PKS 2155$-$304 HVCs were accurately 
determined.  In this sample, where none of the HVCs has accurate values of 
$N$(\ion{H}{1}), the metallicity assumption is useful in restricting parameter 
space and producing reasonable computing times.  Such an assumption does not 
dramatically affect the predicted run of ion column density versus gas density.  
Since these
clouds are optically thin, with presumably subsolar metallicities, all models with
the same product, $Z \times$N(H~I), produce essentially the same results.  We
adopt a typical extragalactic radiation field with log $\phi=4.0$
(photons cm$^{-2}$ s$^{-1}$), where $\phi$ is the normally-incident ionizing
photon flux, and we assume a power-law spectrum, 
$F_{\nu} \propto \nu^{-\alpha}$, with
spectral index $\alpha=1.8$ (Zheng et al.\ 1997; Telfer et al.\ 2002).
Since the intensity of the radiation field is fixed, we ran models for a range
of different ionization parameter, $U$, by varying the hydrogen
number density within the cloud ($U\propto\phi/n_{\rm H}$).

In two sight lines, 3C~273 and PKS~1302$-$102, we only have measurements of
\ion{O}{6}.  With just one ion species, we cannot constrain the ionization
parameter or the gas density.  Thus, we concentrate our efforts on the five
sight lines for which we have measurements of multiple ion species, and
we attempt to determine model parameters that fit these observations.  As
in the models of CSG04, the observed columns of
\ion{O}{6} are far greater than can be explained in a photoionization model
that produces roughly equal amounts of lower ion species.  The ionization
pattern in these HVCs clearly indicates a multiphase structure arising from
several ionization sources.  However, these clouds will be immersed in an
ionizing radiation field, minimally the extragalactic background,
so considering the HVC observations in the context of a photoionization model
is useful for constraining their physical conditions.  For this reason, we
begin by constraining the models with the lowest ionization species observed.
This effectively places an upper limit on the ionization
parameter, or a lower limit on the gas density.  From this information,
we also compute the gas pressure, $P/k=2.3n_{H}T$ (assuming a fully ionized
gas with He$/$H$=0.1$), and the cloud size, $D = N_{H}/n_{H}$.
For the comparisons of models to
observations, we adopt 2$\sigma$ error bars for the measured column densities.
Allowing this flexibility in the comparisons is useful, given the simplicity
of the models and the complexity of the actual situation.  In several cases,
the \ion{Si}{3} and \ion{C}{3} profiles indicate line saturation, and we adopt
the measured value as a lower limit to the column density.

\begin{figure*}
\figurenum{13}
\plotone{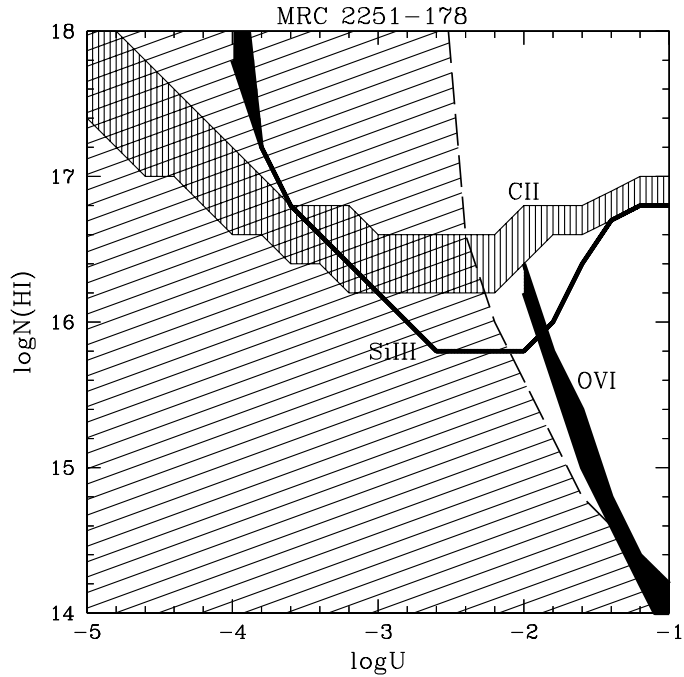}
\caption{Grid of CLOUDY model calculations with model constraints for
the HVC toward MRC 2251$-$178.  The model and its constraint are as in Figure
11.  The constraints on the models based on detections of \ion{Si}{3}
and \ion{O}{6} are represented by the labeled curves.
The constraints based on \ion{C}{2} are represented by the vertical shading.
The large shaded region is the constraint based on the upper limit to
$N$(\ion{N}{5}) (dashed line).  The \ion{C}{2} and \ion{Si}{3} measurements
are consistent with photoionization in an HVC with model parameters in the range
log~$N$(\ion{H}{1}) = 16.2--16.8 and log~$U = -3.6$ to $-3.0$
(log~$n_{H} = -3.5$ to $-2.9$).}
\end{figure*}

We show three grids of photoionization models in Figures 12--14, with 
constraints provided by the observed ion column densities 
indicated by the shaded regions.  The three figures shown, for HVCs
towards HE~0226$-$4110, MRC~2251$-$178, and Mrk~1513, are chosen to
illustrate the range of ionization characteristics within the five
modeled HVCs.  Figure 12 shows the constraints on $N$(\ion{H}{1}) and
log~$U$ for the HVC towards HE~0226$-$4110 based on its ion column
densities.  Since some saturation is present in the \ion{C}{3} and
\ion{Si}{3} profiles, we adopt the measured column densities as lower
limits and plot only the more restrictive \ion{Si}{3} lower limit.
The constraints on the column densities of \ion{C}{2}, \ion{Si}{2},
\ion{C}{3}, and \ion{Si}{3} are consistent with a photoionization
model characterized by log~$N$(\ion{H}{1}) $=17.2$ and an ionization
parameter log~$U = -3.4$ (log~$n_{H} = -3.1$).  These values
correspond to a pressure, $P/k=25$~cm$^{-3}$~K, and cloud size, $D =
5.9$ kpc.  In this model, the predicted $N$(\ion{Si}{4}) is slightly
below and the predicted $N$(\ion{C}{4}) well below the range allowed
by the observations.  However, the excess \ion{Si}{4} and \ion{C}{4}
could arise in a warmer, collisionally ionized layer.  This model
predicts a negligible amount of \ion{O}{6}, indicating once again that
non-photoionizing sources must play a prominent role in ionizing gas
in this HVC.

Figure 13 shows the model constraints based on the ion column density
measurements of the HVC toward MRC~2251$-$178.  Because this sight line was
not observed with the STIS E140M grating, lines of \ion{C}{4} and
\ion{Si}{4} were not measured.  The {\it FUSE}
data are of mediocre quality, so the 2$\sigma$ limits on $N$(\ion{C}{2}) from
$\lambda 1036.34$ do not provide stringent constraints on the
photoionization models.  Based on \ion{C}{2} and \ion{Si}{3}, we find
acceptable models characterized by parameters ranging from
log~$N$(\ion{H}{1})$ =16.2-16.8$ and log~$U = -3.6$ to $-3.0$
(log~$n_{H} = -3.5$ to $-2.9$).  These values correspond to a wide range of
pressures, $P/k = 11.4-37.6$ cm$^{-3}$~K, and cloud sizes, $D =$ 0.9--3.9 kpc.
As in the case of the other sight lines, the observed $N$(\ion{O}{6}) is far
larger than predicted by this model.

\begin{figure*}
\figurenum{14}
\plotone{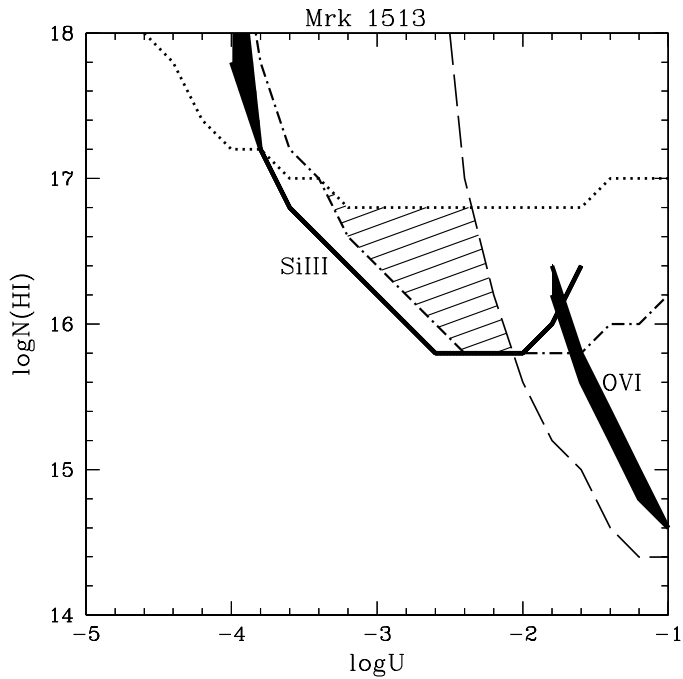}
\caption{Grid of CLOUDY model calculations with model constraints for
the HVC toward MRK 1513.  The model and its constraint are as in Figure
11.  The constraints on the models based on detections of \ion{Si}{3}
and \ion{O}{6} are represented by the labeled curves.  The shaded region
represents the constraints based on the upper limits to $N$(\ion{N}{5})
(dashed line) and \ion{C}{2} (dotted line), and the lower limit to
\ion{C}{3} (dot-dashed line).  The \ion{Si}{3} measurement and the
upper/lower limits are consistent with photoionization in an HVC with
model parameters in the range log~$N$(\ion{H}{1})$=15.8$ and
log~$U=-2.4$ to $-2.2$ (log~$n_{H}=-4.3$ to $-4.1$).}
\end{figure*}

The Mrk 1513 sight line shows different characteristics than the other
sight lines in this study.  Besides \ion{O}{6}, the HVC in this sight
line can be detected in both \ion{Si}{3} and \ion{C}{3}.  The
\ion{C}{3} profile is saturated, and the measured value is taken as
the lower limit.  The \HST STIS G140M data do not cover \ion{C}{2}
$\lambda1334.53$, and the high-velocity \ion{Si}{2} $\lambda1260.42$
profile is contaminated.  Therefore, we are forced to move to the
poorer-quality {\it FUSE} data to search for low ions.  As discussed
in \S~3.4, the {\it FUSE} data for Mrk~1513 are among the worst in the
sample, and we are unable to make a 3$\sigma$ detection of the HVC in
\ion{C}{2} $\lambda1036.34$.  Figure 14 shows the photoionization
model constraints based on the Mrk 1513 HVC column density
measurements.  The upper limits on $N$(\ion{C}{2}) and
$N$(\ion{N}{5}), as well as the lower limit on $N$(\ion{C}{3}),
provide reasonable constraints.  Using the measurement of
$N$(\ion{Si}{3}), we find a best-fit model characterized by
log~$N$(\ion{H}{1})$ = 15.8$ and log~$U = -2.4$ to $-2.2$ (log~$n_{H}
= -4.3$ to $-4.1$), suggesting a less dense medium than the HVCs with
single-ion detections.  These values indicate lower pressures, $P/k =
2.4-3.5$ cm$^{-3}$~K, and larger cloud sizes, $D=28-73$ kpc, than the
HVCs with \ion{C}{2} or \ion{Si}{2} detections.  We note, however,
that this result is highly dependent on the \ion{C}{3} lower limit,
where we have assumed that there is no contamination (the \ion{C}{3}
profile appears different than the \ion{O}{6} and \ion{Si}{3}
profiles).  If the profile were sufficiently contaminated, the model
parameters would be poorly constrained.  For example, if the lower
limit were log~$N$(\ion{C}{3}) $>14.00$ (i.e., 30\% of the profile
optical depth arising from a contaminant line), then the ionization
parameter could be as low as log~$U=-3.0$, and the gas density nearly
an order of magnitude higher.

\begin{deluxetable}{lcccccl}
\tablefontsize{\footnotesize}
\tablecolumns{7}
\tablewidth{0pc}
\tablecaption{HVC MODELING RESULTS \label{t10}}
\tablehead{
\colhead{} & \colhead{log~$N$(\ion{H}{1})} & \colhead{log~$U$} & \colhead{log~$n_H$}
& \colhead{$P/k$\tablenotemark{a}} & \colhead{Size} & \colhead{Ions} \\
\colhead{Sight Line} & \colhead{(cm$^{-2}$)} & \colhead{} & \colhead{(cm$^{-3}$)}
& \colhead{(cm$^{-3}$~K)} & \colhead{(kpc)} & \colhead{for fit\tablenotemark{b}}}
\startdata
HE 0226$-$4110 & 17.2      & $-3.4$  & $-3.1$       & 24.8      &  5.9    & \ion{C}{2
}, \ion{Si}{2} \\
MRC 2251$-$178 & 16.2-16.8 & $-3.6$ to $-3.0$
                                     & $-3.5$ to $-2.9$& 11.4-37.6 & 0.9-3.9 &
                                                        \ion{C}{2}, \ion{Si}{3} \\
MRK 1513       & 15.8      & $-2.4$ to $-2.2$
                                     & $-4.3$ to $-4.1$ & 2.4-3.5   & 28-73   &
                                    \ion{Si}{3}, \ion{C}{3}\tablenotemark{c} \\
PG 0953$+$414  & 16.4      & $-3.3$  & $-3.2$      & 20.6      & 1.5     & \ion{C}{2}
, \ion{Si}{3} \\
TON S180       & 16.6      & $-3.2$  & $-3.3$      & 16.9      & 3.9     & \ion{C}{2}
, \ion{Si}{3}
\enddata
\tablenotetext{a}{For a fully ionized gas with He/H$=0.1$, $P/k=2.3n_{H}T$.}
\tablenotetext{b}{Refers to the measured ions which provide the best
constraints on the fit.}
\tablenotetext{c}{The \ion{C}{3} constraint is a lower limit.}
\end{deluxetable}

The results of the modeling are shown in Table 10.  The models indicate
that HVCs with single-ion detections made in data of {\it FUSE} and
\HST STIS quality (S/N $\geq 5$ per resolution element) have gas densities
of at least log~$n_{H} = -3.5$, pressures $P/k > 10$~cm$^{-3}$~K, and cloud
sizes $D\sim1-6$ kpc.  Since highly ionized HVCs with low ions
(\ion{C}{2}, \ion{Si}{2}) represent 75$\%$ of our sample, the characteristic
gas density in this population seems to be significantly larger than expected
for filaments of the WHIM.  The calculated sizes suggest discrete
clouds, rather than an extended hot medium in the Local Group.  It is
therefore difficult to reconcile these results with the WHIM interpretation
of the high-velocity \ion{O}{6}\ absorbers.

Nicastro \etal (2003) propose that cooled, higher-density regions could
arise in WHIM filaments by compression against virialized structures in the
Local Group.  However, as noted in \S~4.1, the cooling time of the WHIM
is much longer than the Hubble time.   
In addition, there is little direct kinematic evidence for a 
90 \kms drift of the Local Group relative to the WHIM on Mpc distance
scales. Instead, observations (M\'endez \etal\ 2002) show that random 
motions among galaxies out to 550 \kms\ redshift are small, and large
infall velocities only appear on halo scales.   However, even if 
there were large-scale drift of the Local Group, the kinematic similarities 
between the \ion{O}{6} and low-ion
profiles (\ion{C}{2}, \ion{Si}{2}, \ion{C}{3}, \ion{Si}{3}) suggest that
the \ion{O}{6} absorbers detected with \FUSE are more likely associated with
discrete cloud-like objects than with hot WHIM filaments.  We cannot rule out a 
WHIM origin for the two HVCs toward 3C~273 and PKS~1302-102, where the density
cannot be constrained since they are detected only in \ion{O}{6}.

\placetable{t10}

\subsection{Collisional Ionization}

In the previous section, we showed that the ubiquitous detections of low ions
in the highly ionized HVCs are consistent with a scenario where the absorbers
trace photoionized clouds with densities, log~$n_{H}>-3.5$, and pressures
$P/k > 10$ cm$^{-3}$~K.  Although the high ion column densities are
underpredicted in the photoionization models, the kinematic similarities
between the low ion and high ion (\ion{O}{6}, \ion{C}{4}, \ion{Si}{4}) profiles
suggest that these species arise in the same clouds.  The HVCs must be
a multiphase medium, where the low- and high-ion components
trace photoionization and collisionally ionization, respectively.
In this section, we compare the column densities of high ions to various
collisional ionization scenarios, that are physically consistent with the
observed multicomponent HVC structure.  In Table 11, we list column
density ratios involving \ion{O}{6}, \ion{N}{5}, \ion{C}{4}, and \ion{Si}{4}
for the highly ionized HVCs.  We consider only cases where existing data
cover the \ion{C}{4} and \ion{Si}{4} lines.

F04 recently converted several common collisional ionization models into the
case appropriate for the low-metallicity HVC Complex C.  They adopted a
metallicity based on [\ion{O}{1}/\ion{H}{1}] $= -0.79$, as well as a depleted
nitrogen abundance, [N/H] $=-1.85$, a feature common to
low-metallicity absorbers because of a differing nucleosynthetic history
between nitrogen and $\alpha$-process elements (Gibson \etal 2001; Pettini 2004;
Matteucci 2004).  Although the absolute metallicities of these objects are 
unknown, the ion column densities in the PKS~2155$-$304 HVCs are consistent with 
a subsolar abundance pattern.  We therefore use the predictions (F04, Table 7) 
to make a rough comparison to our observed column density ratios.

\begin{deluxetable}{lccc}
\tablefontsize{\footnotesize}
\tablecolumns{4}
\tablewidth{0pc}
\tablecaption{HVC LOGARITHMIC\tablenotemark{a} COLUMN DENSITY RATIOS \label{t11}}
\tablehead{
\colhead{Sight Line} & \colhead{$N$(\ion{C}{4})$/N$(\ion{O}{6})} & \colhead{$N$(\ion{
Si}{4})$/N$(\ion{O}{6})} & \colhead{$N$(\ion{N}{5})$/N$(\ion{O}{6})}}
\startdata
3C 273         & $<-0.88$                & $<-1.36$                & $<-0.71$\\
HE 0226$-$4110 & $-0.21^{+0.07}_{-0.09}$ & $-1.15^{+0.11}_{-0.13}$ & $<-0.51$\\
MRK 509\tablenotemark{b}        &  $0.03^{+0.12}_{-0.09}$ & $-0.64^{+0.17}_{-0.14}$ &
 $<-0.95$\\
PG 0953$+$414  & $-0.64^{+0.09}_{-0.10}$ & $<-1.33$                & $<-0.65$\\
PKS 1302$-$102 & $<-0.91$                & $<-1.29$                & $<-0.56$\\
PKS 2155$-$304\tablenotemark{b}    & $-0.17^{+0.06}_{-0.05}$ & $-1.13^{+0.09}_{-0.09}
$ & $<-0.66$
\enddata
\tablenotetext{a}{The PG~1259+593 sight line through HVC Complex C is
   characterized by logarithmic ratios, log~[N(\ion{C}{4})/N(\ion{O}{6})]
   $ = -0.31^{+0.08}_{-0.09}$, log~[N(\ion{Si}{4})/N(\ion{O}{6})]
   $ = -0.90^{+0.11}_{-0.14}$, and log~[N(\ion{N}{5})/N(\ion{O}{6})]
   $ <-0.38$ (CSG04).}
\tablenotetext{b}{From CSG04.  Although the
reported values are for the full high-velocity absorption range, the HVCs
toward PKS 2155$-$304 and Mrk 509 can be resolved into two components.}
\end{deluxetable}

The upper limits on the column density ratios for the two HVCs detected only 
in \ion{O}{6} (3C~273 and PKS~1302$-$102) can be explained by several collisional 
models, including radiative cooling of hot gas (Edgar \& Chevalier 1986).  The 
3C~273 absorber is a positive-velocity (105--240 \kms) wing to the Galactic 
\ion{O}{6} profile and suggests an outflow from the Galaxy.  The sightline also 
passes near Galactic Radio Loops I and IV
near the North Polar Spur ($\ell = 290\arcdeg, b = 64.4\arcdeg$).
Radiative cooling would be expected if these absorbers trace a Galactic
fountain, although mechanisms appropriate to cloud interfaces
are consistent with the ratios as well.  We find that the 2$\sigma$ error
bars on the high-ion ratios for the low-ion detected HVCs (HE~0226$-$4110,
Mrk~509, PG~0953+414, and PKS~2155$-$304) are roughly consistent with
ionization in turbulent mixing
layers (Slavin, Shull, \& Begelman 1993) or conductive interfaces
(Borkowski \etal 1990).  Such ionization sources would be expected
at the cloud boundary as the HVC interacts with a surrounding hotter medium.
For HVC Complex C, F04 reached a similar conclusion based on data from
the PG~1259+593 sight line.  The column density ratios towards PG~1259+593
listed in Table 11 are quite similar to those observed in the highly ionized
HVCs, suggesting that these objects reside in similar environments.
Complex C is a known halo HVC, characterized by relatively high densities
$n_{H} \gtrsim 0.01$ cm$^{-3}$ (CSG03).  Because the ionization patterns in
Complex C are similar to those of the highly ionized HVCs,
the latter may be low column density analogs to the Galactic halo HVCs
mapped in \ion{H}{1} emission.

\section{Conclusions}

In this paper, we have compiled \HST STIS and \FUSE data for 12 quasar 
sight lines that intercept highly ionized high-velocity clouds.  We 
measured column densities of key ions and investigated the nature of these 
objects through modeling of their observed ionization patterns.  Using these 
results, we arrive at the following conclusions:
\begin{enumerate}

\item  We detect low ions (\ion{C}{2} or \ion{Si}{2}) in 9 of 12 sight
lines, and  doubly-ionized species (\ion{Si}{3} or \ion{C}{3}) in 10 of 
12 sight lines.   These statistics indicate that the highly ionized
HVCs are multiphase structures with significant column densities of low ions, 
in addition to high ions (C~IV, Si~IV, O~VI).  For two absorbers detected 
only in \ion{O}{6}, we cannot rule out a WHIM origin. However, one of them is 
kinematically consistent with hot gas entrained in a Galactic fountain.
The high velocities, O~VI line widths, and O~VI/O~VII column densities are 
consistent with shock-heating and radiative cooling, with post-shock 
temperatures $T_s \approx (1.2 \times 10^6~{\rm K})(V_s/300~{\rm km~s}^{-1})^2$. 

\item Photoionization modeling of the low ions suggests that the clouds with
low-ion detections are characterized by densities log~$n_{H} > -3.5$,
several orders of magnitude larger than that expected for WHIM gas.  The 
kinematic similarity between the low-ion and high-ion profiles suggests that 
the high-velocity \ion{O}{6} is more likely associated with higher density 
clouds in the Galactic halo than with a WHIM filament.

\item Collisional ionization at interfaces with infalling clouds can explain
the presence of high ions in multiphase HVCs.  High-ion line ratios in
the HVCs are similar to those observed in HVC Complex C towards PG 1259+593.
Instead of tracing the WHIM, the highly ionized HVCs may trace Galactic halo 
HVCs, albeit at low column density.

\item  The presence of low ions in the (HVC) O~VI absorbers, together with 
mass constraints described below, suggest that both the O~VI and ($z=0$) O~VII 
absorbers reside in the Galactic halo ($d \leq 100$ kpc).  The O~VII may come 
from a reservoir of up to $10^{10}~M_{\odot}$ gas at $10^{5.5-6.3}$~K
(at 10-30\% solar metallicity), while the O~VI more likely arises in denser, 
cooler interfaces ($10^{5.3-5.7}$~K). 

\end{enumerate}

We close by returning to the three major issues enumerated in \S~1.
These issues hinge on understanding physical properties of these HVCs (density,
temperature, metallicity) and also their distances and locations relative to
the Galaxy and Local Group barycenter.  Our work suggests that
the O~VI HVCs do not reside in low-density gas 1--3 Mpc from the
Milky Way.  The wide range of ionization stages suggests instead that they
reside in the Galactic halo.  The implications of the IGM vs. Galactic halo
debate are profound.

{\bf (1) HVC Mass Problem.}
The location and mass of the high-velocity O~VI absorbers remain controversial, 
since a major reservoir of hot baryons would be required to explain the O~VI 
and ($z=0$ O~VII) absorbers if they were located at Mpc distances.  Local O~VII 
has been seen with {\it Chandra} and {\it XMM/Newton} toward a number of AGN: 
PKS~2155-304, 3C~273, H1821+643, Mrk~421, etc. (Rasmussen \etal 2003; Mathur 
\etal 2003; Fang \etal 2003; Nicastro \etal 2002, 2004; McKernan \etal 2004).  
The typical O~VII column density ($\sim 10^{15}$ cm$^{-2}$) is ten times that 
seen in O~VI ($\sim 10^{14}$ cm$^{-2}$).  This difference is somewhat offset
by the maximum ionization fractions of these ions: 
$f_{\rm OVI} \approx 0.1-0.2$ at $10^{5.45\pm0.1}$~K, versus 
$f_{\rm OVII} \approx 0.95-0.99$ at $10^{5.90\pm0.15}$~K.   
If the O~VI (and O~VII) absorbers are distributed in a spherical shell of WHIM 
(Nicastro \etal 2003), centered around the Local Group at mean radius
$R_{\rm hot} \approx (1~{\rm Mpc})R_{\rm Mpc}$ and typical column density
N$_{\rm OVII} = (10^{15}~{\rm cm}^{-2})N_{15}$, then the total baryonic mass
would need to be,

$  M_{\rm hot} \approx (4 \pi R_{\rm hot}^2) \left[ \frac
       {N_{\rm OVII}\; (1.32m_H)} {(4.9 \times 10^{-4})Z_O \; f_{\rm OVII}}
        \right] $
\begin{equation}
\approx (4 \times 10^{12}~M_{\odot}) \left(
        R_{\rm Mpc}^2 N_{15} Z_{0.1}^{-1} \right) \; .
\end{equation}
We have assumed that (O/H) has a metallicity $Z_O = (0.1~Z_{\odot})Z_{0.1}$
and that the ionization fraction of O~VII is $f_{\rm OVII} = $ 0.40--0.98
at log~$T = 6.0-6.3$.  Similarly large estimates are required for O~VI,
with $f_{\rm OVI} = 0.1-0.2$ at log~$T = 5.4-5.6$.  
This is far too much baryonic mass to be consistent with gravitational models 
of the Local Group (Kahn \& Woltjer 1959; Peebles 1995) which find a total 
mass $\sim 2 \times 10^{12}~M_{\odot}$.  The disagreement becomes much worse 
if one multiplies the above baryon mass by a factor of 6 to account for the
accompanying dark matter [$\Omega_m h^2 = 0.135 \pm 0.009$ and
$\Omega_b h^2 = 0.0224 \pm 0.0009$; Spergel \etal 2003].

{\bf (2) Infalling Gas Clouds in the Halo.}
In hierarchical (CDM) models of Local Group assembly (Kravtsov, Klypin,
\& Hoffman 2002), clumps of gas and dark matter are still falling into
the Galactic halo.  We believe the highly ionized HVCs seen in O~VI and
lower ions may represent bow shocks and wakes produced at the interfaces
of these gas clumps falling into a hot gaseous halo (CSG04).
To produce significant interactions, including ram-pressure stripping,
requires halo gas densities $n_H \geq 10^{-4}$ cm$^{-3}$ (Maloney 2003).
Such densities and pressures do not occur in the WHIM,
and instead require that the HVCs lie within $R < 50$~kpc.
The amount of shock-heating will depend on the Mach number of the infalling
cloud relative to the hot virialized halo or warm photoionized clouds.
Even without shocks, the infalling gas may be subject to viscous
entrainment or turbulent mixing layers.

{\bf (3) Kinematic Signatures of Infall.}
As discussed earlier, a simple kinematic model of infall relative to
the rotating disk and halo can reproduce some of the observed signatures
of the highly ionized HVCs. The HVC velocity pattern (Fig.\ 1) suggests 
that this gas reflects the underlying sense of Galactic rotation,
with strong segregation of red-shifted and blue-shifted absorbers on different
sides of the rotation axis. One should not place too much 
emphasis on the apparent dipole pattern, since we have eliminated sight lines in
the top-right and lower-left regions, which pass through HVC Complexes A, C,
and M, the Magellanic Stream, and other known H~I structures.  Whether the
Local Group might drift relative to the large-scale WHIM
on supercluster length scales depends on the existence of large-scale
flows and peculiar velocities, as well as the nature of gaseous infall
at $D < 50$ kpc.  Recent observations of the local Hubble flow 
(M\'endez \etal 2002) suggest a remarkably quiet flow, with random motions 
of galaxies exhibiting little deviation from a pure Hubble flow  
out to $cz = 550$ \kms.  Small-scale infall within the Local Group is more 
likely, as shown by the relative motion of M31 and the theoretically 
predicted clumpy infall to the Milky Way (Kravtsov \etal 2002).  

To make further progress in understanding the physical state of hot gas
in the Galactic halo will require new information on the distances to
the UV and X-ray absorbers.  An additional challenge will be to relate 
the hot gas (O~VII at $z=0$) with the UV absorbers (O~VI, C~IV, lower ions). 
We noted above the physical difficulties of placing the O~VII absorbers
in the WHIM, at Mpc distances around the Galaxy.  Not only does the
gas mass exceed dynamical limits for the Local Group, but the low
required densities in the WHIM do not allow this hot gas to cool. 
It is far more plausible that the O~VII arises in a reservoir of hot gas 
($10^{6.0-6.3}$~K) in an extended Galactic halo at $d \leq 100$~kpc.  
Equation (1) then suggests a total mass of 
$\sim 10^{10}~M_{\odot}~f_{\rm OVII}^{-1}$ for hot gas ($10^{6.0\pm0.3}$~K) 
with 25\% solar metallicity.  Such hot halos are a natural consequence
of halo virialization.  If $L^*$ galaxies typically have 100-kpc hot halos,
their presence could be detected through O~VII absorption lines in the 
X-ray spectra of bright AGN, at a frequency $dN/dz \approx 4$.

\acknowledgments

Financial support for HVC observations at the University of Colorado comes from
grants provided from the FUSE Project, operated for NASA by the Johns Hopkins
University under NASA contract NAS5-32985.  JMS and MLG also acknowledge
support from theoretical grants from NASA/LTSA (NAG5-7262) and NSF (AST02-06042).

\end{document}